\documentclass[a4paper,english,american]{frontiersSCNS}

\usepackage[T1]{fontenc}
\usepackage{fancyhdr}
\usepackage{xcolor}
\usepackage{array}
\usepackage{prettyref}
\usepackage{float}
\usepackage{multirow}
\usepackage{amsmath}
\usepackage{amssymb}
\usepackage{graphicx}
\usepackage[authoryear]{natbib}
\usepackage{xargs}[2008/03/08]
\usepackage{subscript}

\makeatletter

\pdfpageheight\paperheight
\pdfpagewidth\paperwidth

\providecommand{\tabularnewline}{\\}

\usepackage[T1]{fontenc}

\usepackage{xcolor}
\definecolor{darkblue}{rgb}{0.0,0.0,0.64}

\definecolor{choose_color}{HTML}{B0CFEC}
\definecolor{connect_color}{HTML}{EDAF9B}
\definecolor{rest_color}{HTML}{EBD7A0}

\usepackage{amsmath}

\usepackage[group-separator={,}, group-minimum-digits=4]{siunitx}

\usepackage{prettyref}
\newrefformat{chap}{\hyperref[#1]{Chapter~\ref*{#1}}}
\newrefformat{sec}{\hyperref[#1]{Section~\ref*{#1}}}
\newrefformat{subsec}{\hyperref[#1]{Section~\ref*{#1}}}
\newrefformat{apdx}{\hyperref[#1]{Appendix~\ref*{#1}}}
\newrefformat{lst}{\hyperref[#1]{Listing~\ref*{#1}}}
\newrefformat{fig}{\hyperref[#1]{Figure~\ref*{#1}}}
\newrefformat{tab}{\hyperref[#1]{Table~\ref*{#1}}}
\newrefformat{eqn}{\hyperref[#1]{Equation~\ref*{#1}}}

\usepackage{url,hyperref,lineno,microtype,subcaption}
\usepackage[onehalfspacing]{setspace}

\hypersetup{colorlinks=true, breaklinks=true, linkcolor=darkblue, urlcolor=darkblue, citecolor=darkblue}

\def\keyFont{\fontsize{8}{11}\helveticabold }
\def\firstAuthorLast{Stapmanns {et~al.}}

\def\Authors{Jonas Stapmanns\,$^{1,2,\ast}$, Jan Hahne\,$^{3}$, Moritz Helias\,$^{1,2}$, Matthias Bolten\,$^{3}$, Markus Diesmann\,$^{1,4,5}$ and David Dahmen\,$^{1}$}

\def\Address{
$^1$ Institute of Neuroscience and Medicine (INM-6) and Institute for Advanced Simulation (IAS-6) and JARA Institute Brain Structure Function Relationship (INM-10), J\"ulich Research Centre, J\"ulich, Germany\\
$^2$ Institute for Theoretical Solid State Physics, RWTH Aachen University, Aachen, Germany\\
$^3$ School of Mathematics and Natural Sciences, Bergische Universit{\"a}t Wuppertal, Wuppertal, Germany\\
$^4$ Department of Physics, Faculty 1, RWTH Aachen University, Aachen, Germany\\
$^5$ Department of Psychiatry, Psychotherapy and Psychosomatics, Medical Faculty, RWTH Aachen University, Aachen, Germany
}

\def\corrAuthor{Jonas Stapmanns}

\def\corrEmail{jonas.stapmanns@rwth-aachen.de}

\makeatother

\usepackage{babel}
\begin{document}
\global\long\def\A{\mathbf{A}}%
\global\long\def\A{{\text{A}^{-}}}%
\global\long\def\a{\mathbf{a}}%

\global\long\def\B{\mathbf{B}}%

\global\long\def\b{\mathbf{b}}%

\global\long\def\C{\mathbf{C}}%

\global\long\def\c{\mathbf{c}}%

\global\long\def\Ca{{\text{Ca}^{2+}}}%

\global\long\def\cc{{^{*}}}%

\global\long\def\Cl{{\text{Cl}^{-}}}%

\global\long\def\Cov#1{\text{Cov}\left(#1\right)}%

\global\long\def\d{\mathrm{d}}%

\global\long\def\diff#1#2{{\displaystyle \frac{\text{d}#1}{\text{d}#2}}}%

\global\long\def\D{\mathbf{D}}%

\newcommandx\EW[2][usedefault, addprefix=\global, 1=]{\left\langle #2\right\rangle _{#1}}%

\global\long\def\e{\mathbf{e}}%

\global\long\def\erf{\mathrm{erf}}%

\global\long\def\erfc{\mathrm{erfc}}%

\global\long\def\Em{\mathbf{1}}%

\global\long\def\Ex{\mathcal{E}}%

\global\long\def\diag{\mathrm{diag}}%

\global\long\def\E{\text{E}}%

\global\long\def\ftr{\mathcal{F}}%

\global\long\def\f{\mathbf{f}}%

\global\long\def\Ftr#1#2{\mathfrak{F}[#1](#2)}%

\global\long\def\iFtr#1#2{\mathfrak{F}^{-1}[#1](#2)}%

\global\long\def\h{\mathbf{h}}%

\global\long\def\G{\mathbf{G}}%

\global\long\def\Hz{\:\mathrm{Hz}}%

\global\long\def\In{\mathcal{I}}%

\global\long\def\inp{\text{inp}}%

\global\long\def\I{\text{I}}%

\global\long\def\Int#1#2#3#4{\int\limits _{#3}^{#4} \text{d}#2\ #1}%

\global\long\def\j{\mathbf{j}}%

\global\long\def\J{\mathbf{J}}%

\global\long\def\K{{\text{K}^{+}}}%

\global\long\def\LN{\mathcal{L}_{0}}%

\global\long\def\LO{\mathcal{L}_{1}}%

\global\long\def\Lfp{L_{\mathrm{FP}}}%

\global\long\def\M{\mathbf{M}}%

\global\long\def\m{\mathbf{m}}%

\global\long\def\mus{\:\mu\mathrm{s}}%

\global\long\def\ms{\,\text{ms}}%

\global\long\def\mV{\,\text{mV}}%

\global\long\def\N{\mathbf{N}}%

\global\long\def\n{\mathbf{n}}%

\global\long\def\Na{{\text{Na}^{+}}}%

\global\long\def\nuo{\nu_{0}}%

\global\long\def\nr{n_{r}}%

\global\long\def\OO{\mathbf{O}}%

\global\long\def\P{\mathbf{P}}%

\global\long\def\PV{P(V)}%

\global\long\def\pF{\:\mathrm{pF}}%

\global\long\def\Q{\mathbf{Q}}%

\global\long\def\q{\mathbf{q}}%

\global\long\def\R{\mathbf{R}}%

\global\long\def\r{\mathbf{r}}%

\global\long\def\res{\mathrm{Res}}%

\global\long\def\rest{\text{rest}}%

\global\long\def\s{\mathbf{s}}%

\global\long\def\sps{\text{s}^{-1}}%

\global\long\def\tilPV{\tilde{P}(V)}%

\global\long\def\v{\mathbf{v}}%

\global\long\def\V{\mathbf{V}}%

\global\long\def\Var{\mathrm{Var}}%

\global\long\def\Vth{V_{\theta}}%

\global\long\def\Vr{V_{r}}%

\global\long\def\x{\mathbf{x}}%

\global\long\def\y{\mathbf{y}}%

\global\long\def\T{\mathbf{T}}%

\global\long\def\transp{\mathbf{^{\text{T}}}}%

\global\long\def\uvec{\boldsymbol{u}}%

\global\long\def\U{\mathbf{U}}%

\global\long\def\Var#1{\text{Var}\left(#1\right)}%

\global\long\def\vec#1{\boldsymbol{#1}}%

\global\long\def\w{\boldsymbol{w}}%

\global\long\def\W{\mathbf{W}}%

\global\long\def\X{\mathbf{X}}%

\global\long\def\y{\mathbf{y}}%

\global\long\def\Y{\mathbf{Y}}%

\global\long\def\taue{\tau_{e}}%

\global\long\def\taum{\tau_{\mathrm{m}}}%

\global\long\def\taus{\tau_{\mathrm{s}}}%

\global\long\def\taur{\tau_{\mathrm{r}}}%

\global\long\def\tauM{\tau_{\text{m}}}%

\global\long\def\tauR{\tau_{\text{ref}}}%

\global\long\def\nuext{\nu_{\mathrm{ext}}}%

\global\long\def\defeq{\vcentcolon=}%

\global\long\def\rdefeq{\mathrm{=\vcentcolon}}%

\global\long\def\orderofmagnitude{\sim}%

\global\long\def\unity{\mathds{1}}%

\global\long\def\spikess{\;\mathrm{spikes}/\mathrm{s}}%

\global\long\def\tav#1{\widehat{#1}}%

\global\long\def\av#1{\overline{#1}}%

\global\long\def\dotv{\mathord{\cdot}}%

\global\long\def\Altp{\mathbf{A_{\mathrm{LTP}}}}%

\global\long\def\Altd{\mathbf{A_{\mathrm{LTD}}}}%

\global\long\def\tls{\mathbf{t_{\mathrm{LS}}}}%

\global\long\def\ts{\mathbf{t_{\mathrm{S}}}}%

\global\long\def\ms{\:\mathrm{ms}}%
\global\long\def\mV{\:\mathrm{mV}}%
\global\long\def\gap{\mathrm{gap}}%
\global\long\def\GB{\:\mathrm{GB}}%
\global\long\def\ns{\:\mathrm{nS}}%

\onecolumn\foreignlanguage{english}{}

\noindent 

\noindent 

\noindent 

\firstpage{1}
\author[\firstAuthorLast ]{\Authors}  %This field will be automatically populated 
\address{}                            %This field will be automatically populated 
\correspondance{}                     %This field will be automatically populated
\extraAuth{}                          %If there are more than 1 corresponding author, comment this line and uncomment the next one. 
%\extraAuth{corresponding Author2 \\ Laboratory X2, Institute X2, Department X2, Organization X2, Street X2, City X2 , State XX2 (only USA, Canada and Australia), Zip Code2, X2 Country X2, email2@uni2.edu}
\title[Event-based update of synapses in voltage-based learning rules]{Event-based update of synapses in voltage-based learning rules}%
\phantomsection\addcontentsline{toc}{section}{Title}
\maketitle
\begin{abstract}
Due to the point-like nature of neuronal spiking, efficient neural
network simulators often employ event-based simulation schemes for
synapses. Yet many types of synaptic plasticity rely on the membrane
potential of the postsynaptic cell as a third factor in addition to
pre- and postsynaptic spike times. Synapses therefore require continuous
information to update their strength which a priori necessitates a
continuous update in a time-driven manner. The latter hinders scaling
of simulations to realistic cortical network sizes and relevant time
scales for learning.

Here, we derive two efficient algorithms for archiving postsynaptic
membrane potentials, both compatible with modern simulation engines
based on event-based synapse updates. We theoretically contrast the
two algorithms with a time-driven synapse update scheme to analyze
advantages in terms of memory and computations. We further present
a reference implementation in the spiking neural network simulator
NEST for two prototypical voltage-based plasticity rules: the Clopath
rule and the Urbanczik-Senn rule. For both rules, the two event-based
algorithms significantly outperform the time-driven scheme. Depending
on the amount of data to be stored for plasticity, which heavily differs
between the rules, a strong performance increase can be achieved by
compressing or sampling of information on membrane potentials. Our
results on computational efficiency related to archiving of information
provide guidelines for the design of learning rules in order to make
them practically usable in large-scale networks.

\keyFont{\section{Keywords:}voltage-based plasticity rules, event-based
simulation, spiking neural network simulator, NEST, Clopath rule,
Urbanczik-Senn rule}
\end{abstract}

\newpage
\section{Introduction}

\label{sec:Introduction}

One mechanism for learning in the brain is implemented by changing
the strengths of connections between neurons, known as synaptic plasticity.
Already early on, such plasticity was found to depend on the activity
of the connected neurons. Donald Hebb postulated the principle 'Cells
that fire together, wire together' \citep{Hebb49}. Later on, it was
shown that plasticity is shaped by temporal coordination of activities
even down to the level of individual spikes \citep{Markram97a,Bi98}.
Synaptic plasticity rules for spiking neural networks, such as spike
timing-dependent plasticity (STDP, \citet{Gerstner96}), consequently
employ spike times of pre- and postsynaptic cells to predict the change
in connections.

In recent years, a new class of biologically inspired plasticity rules
has been developed that take into account the membrane potential of
the postsynaptic neuron as an additional factor \citep[for a review, see][]{Mayr10,Gerstner14}.
The rule by \citet{Clopath10_344} can be seen as a prototypical example
for a voltage-based plasticity rule since long-term potentiation of
synapses depends on the presynaptic spike arrival and a filtered version
of the postsynaptic membrane potential. This additional voltage dependence
enables the Clopath rule to describe phenomena that are not covered
by ordinary STDP but can be observed in experimental data, such as
the complex frequency dependence of the synaptic weight changes in
spike pairing experiments \citep{Sjostrom01}. Furthermore it provides
a mechanism for the creation of strong bidirectional connections in
networks, which have been found to be overrepresented in some cortical
areas \citep{Song05_0507}.

Further inspiration for recently proposed plasticity rules originates
from the field of artificial neural networks. These networks showed
great success in the past decade, for example in image or speech recognition
tasks \citep{Hinton06_1527,Krizhevsky12_1097,Hannun14,Lecun2015deep}.
The involved learning paradigms, for example the backpropagation algorithm
\citep{Werbos74,Parker85,LeCun85_599,Rumelhart86_533}, are, however,
often not compatible with biological constraints such as locality
of information for weight updates. To bridge the gap to biology, different
biologically inspired approximations and alternatives to the backpropagation
algorithm have been proposed \citep{Neftci17_324,Sacramento18_8721,Bellec19_738385,Cartiglia20_84}.
A common feature of many of these rules is that weight updates not
only depend on the output activity of pre- and postsynaptic cells,
but also on a third factor, which is a time-continuous signal. A prominent
example of such biologically and functionally inspired rules is the
voltage-based plasticity rule proposed by \citet{Urbanczik14}, where
the difference between somatic and dendritic membrane potential serves
as an error signal that drives learning. This rule, incorporated in
complex microcircuits of multi-compartment neurons, implements local
error-backpropagation \citep{Sacramento18_8721}.

Research on functionally inspired learning rules in biological neural
networks is often led by the requirement to implement a particular
function rather than efficiency. Present studies are therefore primarily
designed to prove that networks with a proposed learning rule minimize
a given objective function. Indeed many learning rules are rather
simple to implement and to test in ad-hoc implementations where at
any point the algorithm has access to all state variables. While the
latter implementations are sufficient for a proof of principle, they
are hard to reuse, reproduce and generalize. In particular, simulations
are restricted to small network sizes, as the simulation code cannot
be straight-forwardly distributed across compute nodes and thus parallelized.
This also limits the simulation speed which is, in particular, problematic
given that successful learning requires simulating networks for long
biological times.

In parallel to the above efforts are long-term developments of simulation
software for biological neural networks \citep[for a review, see][]{Brette07_349}.
Such open-source software, combined with interfaces and simulator-independent
languages \citep{Davison08_11,Djurfeldt10_43,Djurfeldt14}, supports
maintainability and reproducibility, as well as community driven development.
The design of such simulators is primarily led by implementation efficiency.
Code is optimized for neuron and synapse dynamics, with the aim to
upscale simulations to biologically realistic network sizes. A modular
structure of the code facilitates re-use and extensions in functionality.
Therefore, one aim of the community should be the transfer of ad-hoc
proof-of-principle implementations to these well-tested platforms.
Given the differences in design principles behind the exploratory
development of specific models and general-purpose simulation technology,
this transfer is not trivial. In the current study, we show how to
make voltage-based learning rules compatible with spiking neural network
simulators that employ an event-driven update scheme for synapses.

\begin{figure}
  \centering{}\textbf{\includegraphics[width=0.5\textwidth]{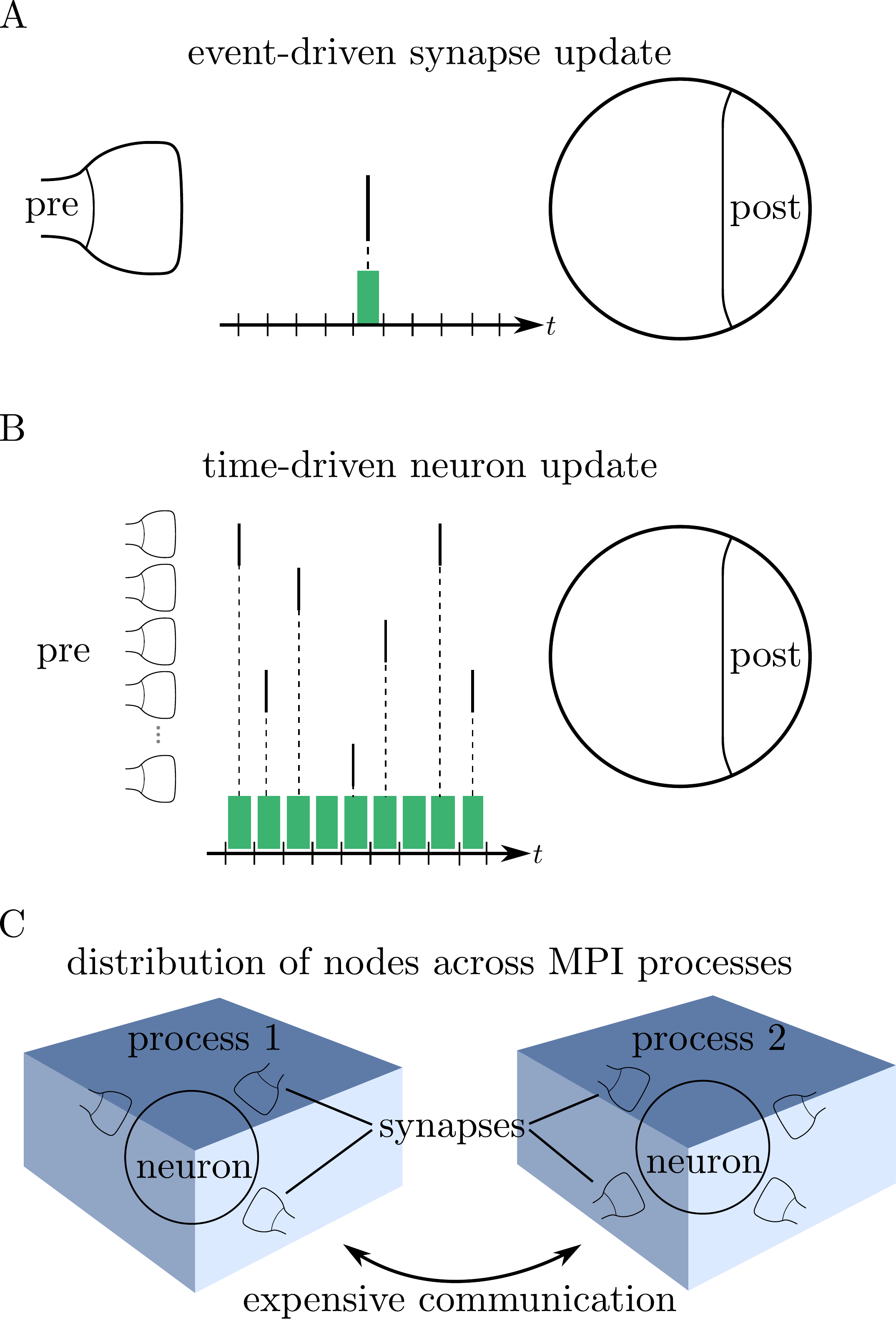}}\caption{\textbf{Update schemes for neurons and synapses. A }A spike crosses
a synapse from the presynaptic (pre) to the postsynaptic (post) neuron.
Since this is a rare event, the synaptic weight is computed only when
a spike is delivered, indicated by the green bar (event-driven update).
\textbf{B} Neurons with a large in-degree receive spikes in rapid
succession which suggests a time-driven update of the neuron's state
in each time step (green bars).\textbf{ C} Since the computation of
the synaptic weights requires information from the postsynaptic neuron,
storing the synapses on the same compute node reduces the amount of
expensive communication between compute processes. \label{fig:Simulation-concepts}}
\end{figure}

Modern network simulators use individual objects for different neurons
and synapses. One common strategy of parallelization is to distribute
these objects across many compute processes \citep{Jordan18_2,Lytton2016_2063}.
Communication between neurons then implies exchange of information
between compute processes. Neurons in the brain primarily communicate
in an event-based fashion via spikes. The duration of these spike
events is on the order milliseconds, which together with typical rates
during physiological brain states of a few spikes per second yields
a coupling that is sparse in time (\prettyref{fig:Simulation-concepts}A).
Spiking simulators emulate this communication by idealizing spikes
as instantaneous events. Thus, in the absence of direct electrical
coupling via gap junctions \citep{Kumar96_381,Hahne15_00022,Jordan20_12},
there is no neuronal interaction in between two spike events such
that the dynamics of neuronal and synaptic state variables can be
evolved independently in time. This led to the development of event-based
simulation schemes, where synapses are only updated in their state
at the times of incoming spikes \citep{Watts94_927,Morrison05a}.
This significantly reduces the amount of function calls to synapse
code and optimizes computational performance in network simulations.
Modern spiking network simulators such as Auryn \citep{Zenke14},
Brian \citep{Stimberg14}, Neuron \citep{Carnevale06}, NEST \citep{Gewaltig_07_11204}
and Nevesim \citep{Pecevski14} therefore employ an event-based update
scheme for synapses. Even though spike events at single synapses are
rare, each single neuron typically receives a large amount of spikes
in rapid succession due to its large number of incoming connections
(in-degree). This suggests a time-driven update of neurons (\prettyref{fig:Simulation-concepts}B).
The resulting hybrid simulation scheme for neurons and synapses \citep{Morrison05a,DHaene14_1055,Krishnan17_75}
is nowadays commonly used across many spiking network simulators \citep[for a review, see][]{Brette07_349}.

An event-based scheme for synapses is perfectly suitable for classical
STDP rules, which only rely on a comparison between the timings of
spike events. Optimizations of simulations including STDP have been
extensively discussed theoretically \citep{Rudolph06_1130,Ros06_2959,Morrison07_1437}
and routinely used in the most common spiking network simulators \citep{Carnevale06,Gewaltig_07_11204,Goodman13,Pecevski14,Zenke14}
as well as in neuromorphic hardware \citep{Serrano13_2,Pfeil13_11,Neftci14_272,Galluppi15_429,Friedmann16_128,Thakur18_891}.
Some STDP variants also include the membrane potential of postsynaptic
cells at the time points of the spike events as a gating variable
\citep{Brader07,Diederich18_1}. At the update, these rules only require
the synapse to know the current value of the postsynaptic membrane
potential in addition to the pre- and postsynaptic spike time. Obtaining
this value from the neuron objects is efficient to implement and already
employed in a range of neuromorphic systems \citep{Serrano13_2,Galluppi15_429,Qiao15_141,Moradi18_106,Cartiglia20_84}.

We here focus on more complex voltage-based learning rules where synapses
continuously require information from the postsynaptic neurons in
order to update their weights \citep{Clopath10_344,Mayr10,Brea13_9565,Yger13_1,Urbanczik14,Albers16_1}.
This a priori breaks the idea behind an event-based update scheme.
Therefore, previous attempts to incorporate such voltage-based plasticity
in spiking network simulators resorted to time-driven synapse updates
for NEST \citep{Jordan20_preprint} and NEURON (see implementation
of Clopath plasticity on ModelDB, \citealt{Hines04_7}). These implementations
therefore only profit from the simulation environment on the level
of the implementation language, but have not been able to exploit
the algorithmic optimizations and speedup of event-based synapse updates.

In this study we present an efficient archiving method for the history
of postsynaptic state variables that allows for an event-based update
of synapses and thus makes complex voltage-based plasticity rules
compatible with state-of-the-art simulation technology for spiking
neural networks. In particular, we derive two event-based algorithms
that store a time-continuous or discontinuous history, respectively.
These algorithms apply to plasticity rules with any dependence on
post-synaptic state variables and therefore cover a large range of
existing models \citep{Brader07,Mayr10,Legenstein11_10787,Yger13_1,Brea13_9565,Qiao15_141,Albers16_1,Sheik16_164,Brea16_1,Diederich18_1,Sacramento18_8721,Cartiglia20_84}.
We theoretically analyze advantages of the two event-driven algorithms
with respect to each other and compare to a straight-forward time-driven
algorithm.

The presented simulation concepts are exemplified and evaluated in
a reference implementation in the open source simulation code NEST
\citep{Gewaltig_07_11204,Nest2180}. The reference implementation
thereby exploits existing functionality of a scalable software platform
which can be used on laptops as well as supercomputers. NEST is employed
by a considerable user community and equipped with an interface to
the programming language Python \citep{Eppler09_12} that is currently
widely used in the field of computational neuroscience \citep{Muller15_11}.
It supports relevant neuron models and connection routines for the
construction of complex networks. Despite this flexibility the simulation
engine shields the researcher from the difficulties of handling a
model description in a distributed setting \citep{Morrison05a,Plesser15_1849}.

To exemplify the general simulation algorithms, we here focus on the
voltage-based plasticity rules by \citet{Clopath10_344} and \citet{Urbanczik14}.
The two rules represent opposing ends of a family of learning rules
in the amount of data required to compute weight updates. The Clopath
rule by design only triggers plasticity in the vicinity of postsynaptic
spike events; storing a history, which is non-continuous in time,
thus becomes beneficial. In contrast, the Urbanczik-Senn rule considers
noisy prediction errors based on postsynaptic membrane voltages and
spikes. Such prediction errors never vanish and therefore always need
to be stored to update the weights, leading to time-continuous histories.
For a given span of biological time, simulations of the Urbanczik-Senn
rule are therefore by design less efficient than those of the Clopath
rule. However, we show that a compression of membrane potential information
reduces this performance gap. Changing the learning rule to include
a sparse sampling of the membrane voltage further increases efficiency
and makes performance comparable to simulations with ordinary STDP.

Our study begins with a specification of the mathematical form of
the learning rules that we consider (\prettyref{subsec:Learning-rules}).
We distinguish between classical STDP (\prettyref{subsec:STDP_formal})
and voltage-based rules (\prettyref{subsec:Voltage-based}) and present
a special case where voltage-based rules can be efficiently implemented
by compressing information on the postsynaptic membrane potential.
We then introduce the Clopath and the Urbanczik-Senn rule as two examples
of voltage-based plasticity (\prettyref{subsec:Example_Cl} and \prettyref{subsec:Example_US}).
In \prettyref{sec:Results-1} we first contrast time- and event-driven
schemes for updating synapses with voltage-based plasticity (\prettyref{subsec:Simulation-concepts}).
Subsequently, we detail a reference implementation of the algorithms
in NEST (\prettyref{subsec:Reference-implementation}) and use this
to reproduce results from the literature (\prettyref{subsec:Reproduction-of-results}).
After that, we examine the performance of the reference implementation
for the Clopath and the Urbanczik-Senn rule (\prettyref{subsec:Performance}).
Conclusions from the implementation of the two rules are drawn in
\prettyref{subsec:Conclusions}, followed by a general Discussion
in \prettyref{sec:Discussion}. The technology described in the present
article is available in the 2.20.1 release of the simulation software
NEST as open source. The conceptual and algorithmic work is a module
in our long-term collaborative project to provide the technology for
neural systems simulations \citep{Gewaltig_07_11204}.

\section{Materials and Methods}

\label{sec:Material-and-methods}

\subsection{General structure of learning rules\label{subsec:Learning-rules}}

\begin{figure}
  \centering{}\includegraphics[width=0.4\linewidth]{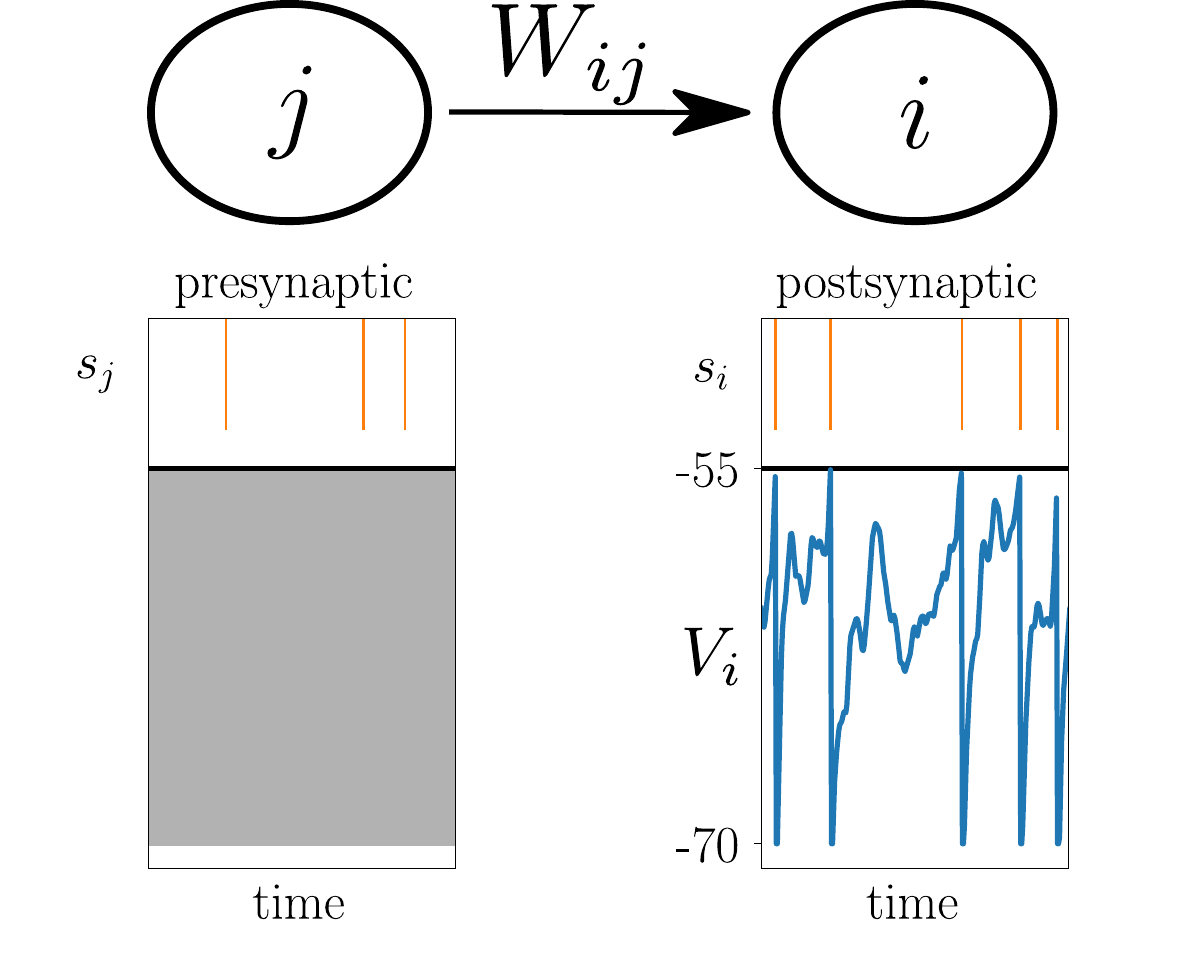}\caption{\textbf{Voltage-based plasticity rules.} The change $\Delta W_{ij}$
in synaptic strength between presynaptic neuron $j$ and postsynaptic
neuron $i$ depends on the presynaptic spike train $s_{j}$, the postsynaptic
spike train $s_{i}$ and the postsynaptic membrane potential $V_{i}$.
\label{fig:info}}
\end{figure}

The focus of this study are plasticity models of the general form

\begin{align}
\frac{dW_{ij}(t)}{dt} & =F(W_{ij}(t),s_{i}^{*}(t),s_{j}^{*}(t),V_{i}^{*}(t))\label{eq:general_plasticity}
\end{align}
where the change $\frac{dW_{ij}(t)}{dt}$ of the synaptic weight $W_{ij}$
between the presynaptic neuron $j$ and postsynaptic neuron $i$ is
given by a function $F$ that potentially depends on the current synaptic
weight $W_{ij}(t)$, as well as on $s_{i}^{*}(t),s_{j}^{*}(t),V_{i}^{*}(t)$
which are causal functionals of the postsynaptic spike train $s_{i}$,
the presynaptic spike train $s_{j}$, and the postsynaptic membrane
potential $V_{i}$, respectively (\prettyref{fig:info}). Causal
functional here refers to $s_{i}^{\ast}(t)$ potentially depending
on all past values $s_{i}(t^{\prime}\le t)$; likewise $V^{\ast}(t)$
depends on $V(t^{\prime}\le t)$. Note that for simplicity of the
notation, we only show one function $F$ on the right hand side of
\prettyref{eq:general_plasticity}, while generally there could be
a sum of multiple functions or functionals $F_{\alpha}$, where each
one depends on spike trains and membrane potentials in a different
manner. Note also that $F$ mixes information of pre- and postsynaptic
neurons, while the functionals denoted by $*$ only need to take into
account information of either the pre- or postsynaptic neuron. In
cases where $F$ is a functional, i.e. where $F$ depends on the
whole time course of its arguments, it can take into account an additional
joint history dependence on $s_{i}^{*},s_{j}^{*}$ and $V_{i}^{*}$.
A special case, the Urbanczik-Senn learning rule, is discussed further
below.

One can formally integrate \prettyref{eq:general_plasticity} to obtain
the weight change between two arbitrary time points $t$ and $T$
\begin{equation}
\Delta W_{ij}(t,T)=\int_{t}^{T}dt^{\prime}F(W_{ij}(t^{\prime}),s_{i}^{*}(t^{\prime}),s_{j}^{*}(t^{\prime}),V_{i}^{*}(t^{\prime})).\label{eq:int_plasticity}
\end{equation}

\subsection{Spike-timing dependent plasticity\label{subsec:STDP_formal}}

In general, the integral on the right hand side of the equation cannot
be calculated analytically. There is, however, a notable exception,
which is the model of spike-timing dependent plasticity (STDP). This
model is a form of Hebbian plasticity that relies on the exact spike
times of pre- and postsynaptic neurons and ignores any effect of the
postsynaptic membrane potential. The dependence on the exact spike
times becomes apparent by the fact that either the pre- or postsynaptic
spike functional is the spike train itself, for example
\begin{equation}
s_{i}^{*}(t)=s_{i}(t)=\sum_{k}\delta(t-t_{i}^{k}),
\end{equation}
where $t_{i}^{k}$ is the $k$-th spike of the $i$-th neuron. This
yields a plasticity rule that reads \citep{Morrison08_459}
\begin{equation}
\frac{dW_{ij}(t)}{dt}=-f_{-}(W_{ij}(t))s_{-,i}^{*}(t)s_{j}(t)+f_{+}(W_{ij}(t))s_{+,j}^{*}(t)s_{i}(t)\label{eq:STDP}
\end{equation}
with functions $f_{\pm}$ that model the weight dependence, and functionals
$s_{\pm}^{*}(t)=(\kappa_{\pm}\ast s)(t)$ given as convolutions of
spike trains with kernels $\kappa_{\pm}$, which in the classical
STDP rule correspond to one-sided exponential decays. The appearance
of the raw spike trains (delta distributions) in the differential
equation of the STDP model renders the integration of the ODE trivial
\begin{equation}
\Delta W_{ij}(t,T)=-\sum_{\mathrm{spikes\,}k}f_{-}(W_{ij}(t_{j}^{k}))\kappa_{-,i}(t_{j}^{k})+\sum_{\mathrm{spikes\,}l}f_{+}(W_{ij}(t_{i}^{l}))\kappa_{+,j}(t_{i}^{l}),\label{eq:int_STDP}
\end{equation}
where $t_{j}^{k},t_{i}^{l}\in[t,T]$. An update of the synaptic weight
between any two time points only requires knowledge of the weight
and spike functionals at the timing of the pre- and postsynaptic spikes.

For models that do not solely rely on exact spike times, but for example
on filtered versions of the spike trains, much more information is
needed in order to calculate a weight update $\Delta W_{ij}(t,T)$
between any two time points. This makes the computation more involved:
the synapse needs all values of $W_{ij}(t^{\prime}),s_{i}^{*}(t^{\prime}),s_{j}^{*}(t^{\prime}),V_{i}^{*}(t^{\prime})$
for $t^{\prime}\in[t,T]$ to update its weight. The remainder of this
study describes different approaches to this problem and their advantages
and disadvantages.

\subsection{Voltage-based plasticity\label{subsec:Voltage-based}}

In a time-driven neuron update, the membrane potential in many simulators
is computed at each simulation step $t^{\alpha}=\alpha\cdot h$, where
$h$ is the simulation step size and $\alpha\in\mathbb{N}$. For plasticity
models that rely on the membrane potential, the time discretization
of \prettyref{eq:int_plasticity} therefore yields

\begin{align}
\Delta W_{ij}(t,T) & =\sum_{\mathrm{steps\,}\alpha}\Delta W_{ij}(t^{\alpha},t^{\alpha+1}),\label{eq:int_plasticity-1}\\
\Delta W_{ij}(t^{\alpha},t^{\alpha+1}) & =\int_{t^{\alpha}}^{t^{\alpha+1}}dt^{\prime}F(W_{ij}(t^{\prime}),s_{i}^{*}(t^{\prime}),s_{j}^{*}(t^{\prime}),V_{i}^{*}(t^{\alpha})).\label{eq:DeltaW}
\end{align}
which, in comparison to the small sum over spikes in the STDP rule
\prettyref{eq:int_STDP}, contains a large sum over all time steps
$t^{\alpha}$ in between time points $t$ and $T$. As the membrane
potential is only known at time points $t^{\prime}=t^{\alpha},$ it
enters \prettyref{eq:DeltaW} in a piecewise constant manner -- hence
the argument $V(t^{\alpha})$. The synapse therefore predominantly
needs information of the postsynaptic neuron in order to update its
weight. Thus, in a distributed simulation framework, where neurons
are split across multiple compute processes, it is beneficial to store
the synapses at the site of the postsynaptic neurons in order to reduce
communication (\prettyref{fig:Simulation-concepts}C). This confirms
the earlier design decision of \citet{Morrison05a} who place synapses
at the site of the postsynaptic neuron to reduce the amount of data
communicated by the presynaptic site.

If weight changes $\Delta W_{ij}$ depend on the synaptic weight themselves,
then \prettyref{eq:DeltaW} cannot be used in practice as intermediate
weights $W_{ij}(t^{\prime})$ for $t^{\alpha}<t^{\prime}<t^{\alpha+1}$
are not known. In this scenario, weight changes have to be calculated
on the simulation grid with $W_{ij}(t^{\prime})\rightarrow W_{ij}(t^{\alpha})$
in case of a forward Euler scheme, or $W_{ij}(t^{\prime})\rightarrow W_{ij}(t^{\alpha+1})$
in case of a backward Euler scheme. In the following we, for simplicity,
stick to the forward Euler setting and arrive at the core computation
for voltage-based plasticity rules
\begin{equation}
\Delta W_{ij}(t^{\alpha},t^{\alpha+1})=\int_{t^{\alpha}}^{t^{\alpha+1}}dt^{\prime}F(W_{ij}(t^{\alpha}),s_{i}^{*}(t^{\prime}),s_{j}^{*}(t^{\prime}),V_{i}^{*}(t^{\alpha})).\label{eq:core}
\end{equation}
Given that $s_{i}$ and $s_{j}$ are spike trains, the functionals
$s_{i}^{*}$ and $s_{j}^{*}$ are obtained trivially from the kernels
of their corresponding Volterra expansions. If $F$ in addition does
not depend on $s_{i}^{*}$ and $s_{j}^{*}$ in a too complicated manner,
which is usually the case (see examples below), the integral in \prettyref{eq:core}
can be calculated analytically.

\subsubsection{Compression of postsynaptic information\label{subsec:Compression-methods}}

The major operation of the plasticity scheme in terms of frequency
and complexity is the computation of infinitesimal weight changes
$\Delta W_{ij}(t^{\alpha},t^{\alpha+1})$. Since the presynaptic spike
train $s_{j}^{*}$ enters $F$ in \prettyref{eq:core}, the same postsynaptic
information on $s_{i}^{*}$ and $V_{i}^{*}$ is used many times for
very similar computations: the membrane potential trace of each neuron
is effectively integrated many times. Is there a way to employ the
result of the computation $\Delta W_{ij}(t^{\alpha},t^{\alpha+1})$
for neuron $j$ for the computations $\Delta W_{ik}(t^{\alpha},t^{\alpha+1})$
for other neurons $k\neq j$? In a simple setting, where $F$ factorizes
into $F(W_{ij}(t),s_{i}^{*}(t),s_{j}^{*}(t),V_{i}^{*}(t))=s_{j}^{*}(t)\,G\left(s_{i}^{*}(t),V_{i}^{*}(t)\right)$
with $s_{j}^{*}\left(t\right)=\left(\kappa\ast s_{j}\right)\left(t\right)$
and
\begin{equation}
\kappa\left(t\right)=H\left(t\right)\frac{1}{\tau}e^{-\frac{t}{\tau}},\label{eq:kappa_conv_kernel}
\end{equation}
defined via the Heaviside step function $H\left(x\right)$, we can
make use of the property
\begin{equation}
s_{j}^{*}(t)=(s_{j}^{*}(t_{LS})+\tau^{-1})\,e^{-(t-t_{LS})/\tau},
\end{equation}
where $t>t_{LS}$ and $t_{LS}$ denotes the last spike time of the
presynaptic neuron. In this case the weight update in between two
spike events factorizes
\begin{align}
\Delta W_{ij}(t_{LS},t_{\mathrm{S}}) & =\underbrace{\left(s_{j}^{*}(t_{LS})+{\color{brown}\tau^{-1}}\right)}_{=:\bar{x}_{j}(t_{LS})}\underbrace{\int_{t_{LS}}^{t_{\mathrm{S}}}dt^{\prime}e^{-(t^{\prime}-t_{LS})/\tau}\,G\left(s_{i}^{*}(t^{\prime}),V_{i}^{*}(t^{\prime})\right)}_{=:\Delta W_{i}(t_{LS},t_{\mathrm{S}})}\label{eq:Delta W_i}
\end{align}
where the latter integral $\Delta W_{i}(t_{LS},t_{\mathrm{S}})$ is
independent of the presynaptic spike train $s_{j}^{*}$. Moreover,
$\Delta W_{i}$ depends on $t_{LS}$ only via an exponential prefactor.
Thus, an integral $\Delta W_{i}(t_{1},t_{\mathrm{2}})$ over an arbitrary
time interval $t_{\mathrel{LS}}\leq t_{1}<t_{2}\leq t_{S}$ which
is completely independent of any presynaptic information, can be used
as a part of the whole integral $\Delta W_{i}(t_{LS},t_{\mathrm{S}})$
since it can be decomposed as
\[
\Delta W_{i}(t_{LS},t_{\mathrm{S}})=\Delta W_{i}(t_{LS},t_{1})+e^{-\frac{t_{1}-t_{\mathrm{LS}}}{\tau}}\Delta W_{i}(t_{1},t_{\mathrm{2}})+e^{-\frac{t_{2}-t_{\mathrm{LS}}}{\tau}}\Delta W_{i}(t_{2},t_{S}).
\]
Therefore, whenever an integral of the postsynaptic quantities $s_{i}^{*}$
and $V_{i}^{*}$ is computed, it can be used to advance the weight
update of all incoming connections and the integration only needs
to be performed once. To account for the generally different last
spike times $t_{\mathrm{LS}}$ of the incoming connections, the postsynaptic
neuron stores the different $\Delta W_{i}\left(t_{LS},t\right)$ in
a so-called \textit{compressed history}. At the time of an incoming
spike event, $\Delta W_{i}(t_{LS},t_{\mathrm{S}})$ can be read out
by the synapse for the correct $t_{LS}$ of that synapse and be combined
with the stored presynaptic spike trace $s_{j}^{*}$.

\subsection{Example 1: Clopath plasticity\label{subsec:Example_Cl}}

The Clopath rule \citep{Clopath10_344} was designed as a voltage-based
STDP rule that accounts for nonlinear effects of spike frequency on
weight changes which had been previously observed in experiments \citep{Sjostrom01}.
It does so by using the evolution of the postsynaptic membrane voltage
around postsynaptic spike events instead of the postsynaptic spikes
themselves. This requires a neuron model that takes into account features
of membrane potential excursions near spike events, such as modified
adaptive exponential integrate-and-fire (aeif) model neurons that
are used in the original publication (\citet{Clopath10_344}, see
\prettyref{subsec:Voltage-clamping}) or Hodgkin-Huxley (hh) neurons
that are used in a NEURON reference implementation on ModelDB \citep{Hines04_7}.

\begin{figure}
\begin{centering}
  \includegraphics[width=0.8\textwidth]{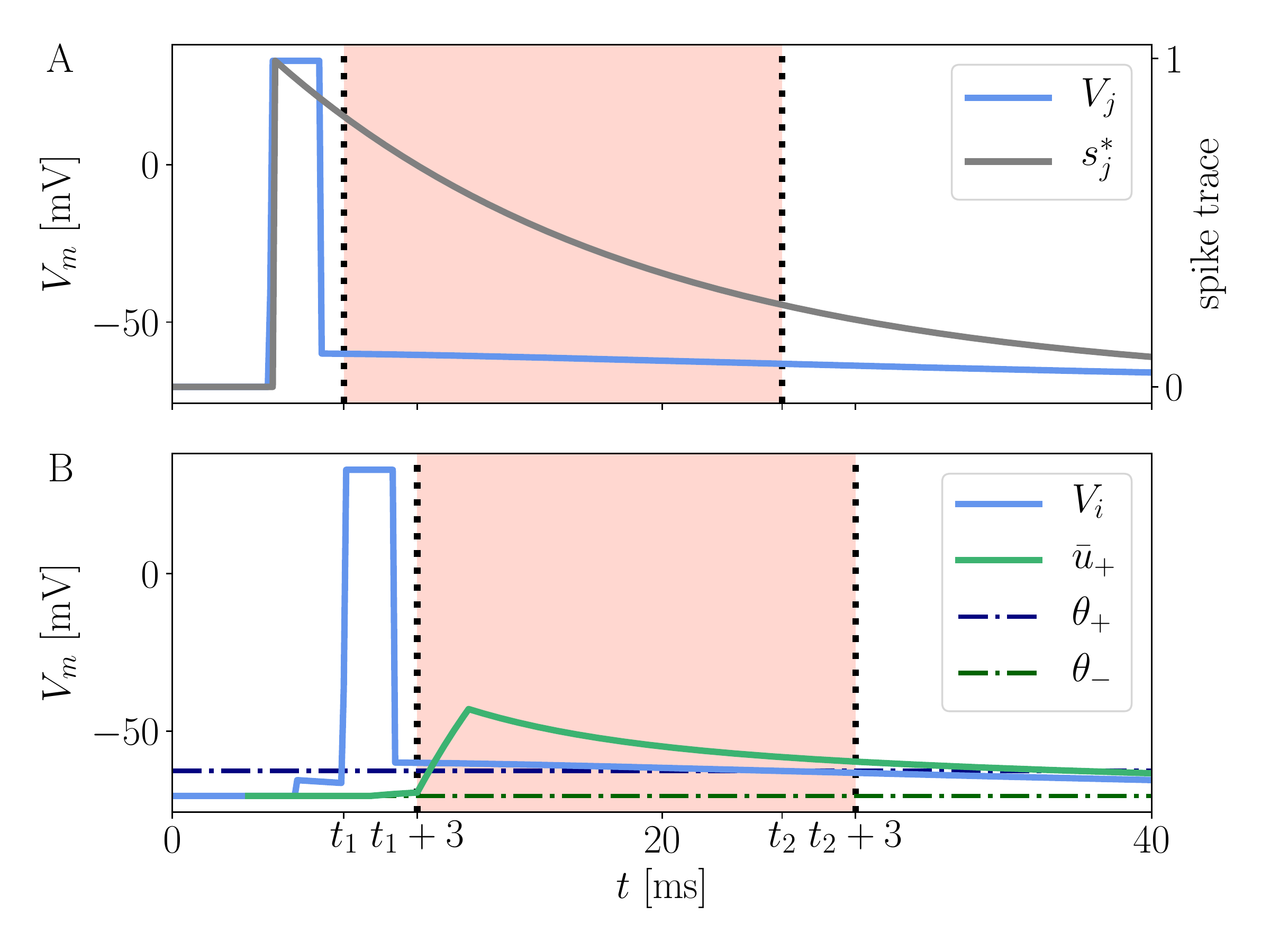}
\par\end{centering}
\caption{\textbf{Illustration of LTP contribution to the Clopath rule}. A presynaptic
neuron (panel \textbf{A}) and a postsynaptic neuron (panel \textbf{B})
emit a spike at $t_{\mathrm{sp,pre}}=4\,\mathrm{ms}$ and $t_{\mathrm{sp,post}}=6\,\mathrm{ms}$,
respectively. The presynaptic spike elicits a trace $s_{j}^{*}$ (gray)
at the synapse. The excursion of the postsynaptic membrane potential
$V_{i}$ (panel B, blue) elevates the low-pass filtered potential
$\bar{u}_{+}$ (green) so that both $V_{i}$ and $\bar{u}_{+}$ exceed
the respective thresholds $\theta_{+}$ (dash-dotted, dark blue) and
$\theta_{-}$ (dash-dotted, dark green), cf. \prettyref{eq:clopath_ltp},
between $t_{1}$ and $t_{2}$. Only within this period, shifted by
$d_{s}=3\,\mathrm{ms}$, which is for times $t_{1}+3\,\mathrm{ms}<t<t_{2}+3\,\mathrm{ms}$
(panel B, red background), see \prettyref{subsec:details_cl} for
details, the LTP of the synaptic weight is non-vanishing because of
the threshold-linear functions in \ref{eq:clopath_ltp}. The shift
by $d_{s}=3\,\mathrm{ms}$ does not apply to the spike trace (panel
A, red background). The rectangular shape of the spikes is achieved
by a clamping of the membrane potential to $V_{\mathrm{clamp}}=33\,\mathrm{mV}$
for a period of $t_{\mathrm{clamp}}=2\,\mathrm{ms}$. \label{fig:Illustration-clopath}}
\end{figure}

The plasticity rule is of the general form \prettyref{eq:general_plasticity}
with a sum of two different functions $F_{\alpha}$ on the right hand
side. It treats long-term depression (LTD) and potentiation (LTP)
of the synaptic weight in the two terms $F_{\mathrm{LTD}}$ and $F_{\mathrm{LTP}}$,
with

\begin{align}
F_{\mathrm{LTD}}\left(s_{j}(t),V_{i,\mathrm{LTD}}^{*}(t)\right) & =-A_{\mathrm{LTD}}\,\,s_{j}(t)\,V_{i,\mathrm{LTD}}^{*}(t)\label{eq:clopath_ltd}\\
\mathrm{with\,\,\,}V_{i,\mathrm{LTD}}^{*} & =(\bar{u}_{-}\,-\,\theta_{-})_{+}\,,\nonumber \\
\bar{u}_{-}(t) & =(\kappa_{-}\ast V_{i})(t-d_{s})\nonumber 
\end{align}
and
\begin{align}
F_{\mathrm{LTP}}\left(s_{j}^{*}(t),V_{i,\mathrm{LTP}}^{*}(t)\right) & =A_{\mathrm{LTP}}\,\,s_{j}^{*}(t)\,V_{i,\mathrm{LTP}}^{*}(t)\label{eq:clopath_ltp}\\
\mathrm{with\,\,\,\,\,\,\,\,\,}s_{j}^{*} & =\kappa_{s}\ast s_{j}\,,\nonumber \\
V_{i,\mathrm{LTP}}^{*} & =(\bar{u}_{+}\,-\,\theta_{-})_{+}(V_{i}\,-\,\theta_{+})_{+}\,,\nonumber \\
\bar{u}_{+}(t) & =(\kappa_{+}\ast V_{i})(t-d_{s})\,.\nonumber 
\end{align}
Here $\left(x-x_{0}\right)_{+}=H(x-x_{0})\left(x-x_{0}\right)$ is
the threshold-linear function and $H\left(x\right)$ is the Heaviside
step function. $A_{\mathrm{LTD}}$ and $A_{\mathrm{LTP}}$ are prefactors
controlling the relative strength of the two contributions. $\kappa_{\pm}$
are exponential kernels of the form \prettyref{eq:kappa_conv_kernel},
which are applied to the postsynaptic membrane potential, and $\kappa_{s}$
is an exponential kernel applied to the presynaptic spike train. The
time-independent parameters $\theta_{\pm}$ serve as thresholds below
which the (low-pass filtered) membrane potential does not cause any
weight change (\prettyref{fig:Illustration-clopath}). Note that $A_{\mathrm{LTP}}$
can also depend on the membrane potential. This case is described
in Appendix \prettyref{subsec:Implementation-ubarbar}.

In a reference implementation of the Clopath rule by C. Clopath and
B. Torben-Nielsen available on ModelDB \citep{Hines04_7}, there is
a subtle detail not explicitly addressed in the original journal article.
In their implementation the authors introduce an additional delay
$d_{s}$ between the convolved version of the membrane potentials
$\bar{u}_{\pm}$ and the bare one (cf. parameter $d_{s}$ in \prettyref{eq:clopath_ltd}
and \prettyref{eq:clopath_ltp} ). The convolved potentials are shifted
backwards in time by the duration of a spike $d_{s}$ (see Tables
S1 and S3 in supplement). As a result, the detailed shape of the
excursion of the membrane potential during a spike of the postsynaptic
neuron does not affect the LTP directly but only indirectly via the
low-pass filtered version $\bar{u}_{+}$, see green background in
\prettyref{fig:Illustration-clopath}B. Incorporating this time shift
in $\bar{u}_{\pm}$ is essential to reproduce the results from \citet{Clopath10_344}
on spike-pairing experiments.

The depression term $F_{\mathrm{LTD}}$ depends on the unfiltered
spike train $s_{j}$. It can thus be treated analogous to ordinary
STDP rules (cf. \prettyref{eq:STDP}ff). In particular, $V_{i,\mathrm{LTD}}^{*}$
only needs to be available for time points of presynaptic spikes (potentially
taking into account additional delays of the connection). The potentiation
term $F_{\mathrm{LTP}}$, however, depends on the filtered spike train
$s_{j}^{*}$; $V_{i,\mathrm{LTP}}^{*}$ consequently needs to be known
also for times in between spike events.

\subsection{Example 2: Urbanczik-Senn plasticity\label{subsec:Example_US}}

The Urbanczik-Senn rule \citep{Urbanczik14} applies to synapses
that connect to dendrites of multicompartment model neurons. The main
idea of this learning rule is to adjust the weights of dendritic synapses
such that the dendrite can predict the firing rate of the soma. The
dendrite expects the firing rate to be high when the dendrite's membrane
potential is elevated due to many incoming spikes at the dendrite,
and to be low if there are only a few incoming spikes. Thus, for this
prediction to be true, synapses that transmit a spike towards the
dendrite while the firing rate of the soma is low are depressed and
those that provide input while the soma's firing rate is high are
facilitated. Learning can be triggered by applying a teacher signal
to the neuron via somatic synapses such that the actual somatic firing
deviates from the dendritic prediction.

The plasticity rule is again of the general form \prettyref{eq:general_plasticity},
with a functional $F$ on the right hand side that reads

\begin{align}
F[s_{j}^{*},V_{i}^{*}]= & \eta\,\kappa\ast\left(V_{i}^{*}\,s_{j}^{*}\right)\label{eq:F_US}\\
\mathrm{with\,\,\,}V_{i}^{*} & =\left(s_{i}-\phi(V_{i})\right)\,h\left(V_{i}\right),\label{eq:prediction_error}\\
s_{j}^{*} & =\kappa_{s}\ast s_{j}.\nonumber 
\end{align}
with exponential filter kernels $\kappa$ and $\kappa_{s}$ and nonlinearities
$\phi$ and $h$. Note that $F$ depends on the postsynaptic spike
train $s_{i}$ via $V_{i}^{*}$. The latter can be interpreted as
a prediction error, which never vanishes as spikes $s_{i}$ (point
process) are compared against a rate prediction $\phi(V_{i})$ (continuous
signal).

In order to solve \prettyref{eq:general_plasticity}, we need to integrate
over $F[s_{j}^{*},V_{i}^{*}]$ (cf. \prettyref{eq:int_plasticity}).
Writing down the convolution with $\kappa$ explicitly, we obtain
\begin{align}
\Delta W_{ij}(t,T) & =\int_{t}^{T}dt^{\prime}\,F[s_{j}^{*},V_{i}^{*}](t^{\prime})\nonumber \\
 & =\int_{t}^{T}dt^{\prime}\,\eta\int_{0}^{t^{\prime}}dt^{\prime\prime}\kappa\left(t^{\prime}-t^{\prime\prime}\right)V_{i}^{*}\left(t^{\prime\prime}\right)s_{j}^{*}\left(t^{\prime\prime}\right).\label{eq:us_weight_bad}
\end{align}
A straight forward implementation of this expression is inefficient
in terms of memory usage and computations because of the two nested
integrals. However, since the kernels $\kappa$ and $\kappa_{s}$
are exponentials, one can perform one of the integrations analytically
(see appendix \prettyref{subsec:Analytical-integration-in} for a
derivation) to rewrite the weight change as
\begin{alignat}{1}
\Delta W_{ij}(t,T) & =\eta\left[I_{1}\left(t,T\right)-I_{2}\left(t,T\right)+I_{2}\left(0,t\right)\left(1-e^{-\frac{T-t}{\tau_{\kappa}}}\right)\right],\label{eq:us_weight_change}\\
\mathrm{with}\,\,\,I_{1}\left(a,b\right) & =\int_{a}^{b}dt\,V_{i}^{*}\left(t\right)s_{j}^{*}\left(t\right),\nonumber \\
I_{2}\left(a,b\right) & =\int_{a}^{b}dt\,e^{-\frac{b-t}{\tau_{\kappa}}}V_{i}^{*}\left(t\right)s_{j}^{*}\left(t\right)\,,\nonumber 
\end{alignat}
which is in line with the general formulation discussed in \prettyref{subsec:Voltage-based}.

\section{Results}

\label{sec:Results-1}

In the following, we first discuss time- and event-driven update schemes
for synapses with voltage-based plasticity. Then we present a reference
implementation for the Clopath rule \citep{Clopath10_344} and the
Urbanczik-Senn rule \citep{Urbanczik14} in the spiking network simulator
NEST \citep{Nest2180}. Finally, we show that these implementations
reproduce results of the original works and we assess their simulation
performance on a distributed computing architecture.

\subsection{Time-driven vs event-driven update scheme for synapses with voltage-based
plasticity\label{subsec:Simulation-concepts}}

\begin{figure}
  \centering{}\includegraphics[width=1\linewidth]{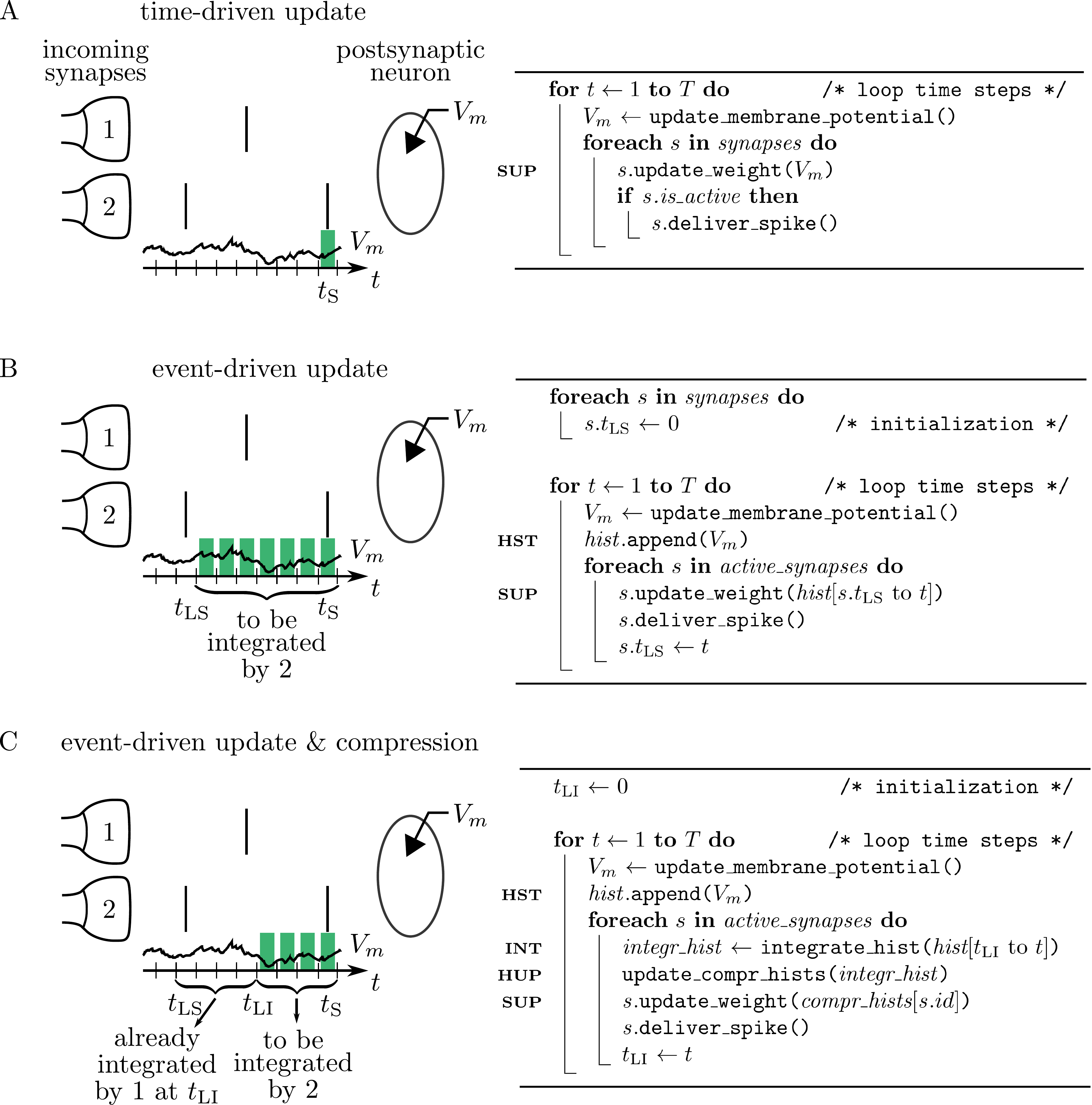}\caption{\textbf{Simulation concepts. }Left: illustration of processing the
postsynaptic voltage trace $V_{m}\left(t\right)$ for three simulation
concepts. Two incoming synapses (1 and 2) transmit spikes (black,
vertical bars) to the postsynaptic neuron. Depending on the algorithm,
a different number of past membrane potentials has to be stored (green
blocks) so that synapse 2 can update its weight when it delivers the
spike at time $t_{s}$. Right: corresponding pseudocodes. \textbf{A}
In the time-driven update scheme the synaptic weight change is evaluated
in every time step of the simulation for all the synapses. This requires
only the latest value of the membrane potential to be accessible by
the synapse to update its weight at $t_{S}$ (see line marked SUP
in pseudocode). \textbf{B} In the event-driven update scheme the computation
of the synaptic weight change is performed only if a spike crosses
the synapse. Therefore, storage of the time trace of $V_{m}$ (see
HST in code) from the last spike delivered by synapse 2 at $t_{\mathrm{LS}}$
up to the current time step $t_{S}$ is needed. \textbf{C} In the
compressed event-driven update scheme synapse 2 uses the time trace
of $V_{m}$ integrated from the last incoming spike at $t_{\mathrm{LI}}$
up to the current time step $t_{S}$ (see INT in code) to complete
its weight update (see SUP in code) and also to advance that of synapse
1. The preceding part of $V_{m}$ from $t_{\mathrm{LS}}$ to $t_{\mathrm{LI}}$
was already integrated and applied to all incoming synapses (see HUP
in code) by synapse 1 when it delivered the spike at $t_{\mathrm{LI}}$.\label{fig:Simulation-concepts-1}}
\end{figure}

\begin{table}
\begin{centering}
\begin{tabular}{|c|c|c|c|}
\hline 
 & time-driven & event-driven & \multicolumn{1}{c|}{event-driven \& compression}\tabularnewline
\hline 
history length $L$ & $1$ & $I$ & \multirow{1}{*}{$i$}\tabularnewline
\hline 
synapse function calls $M$ & $K\cdot T$ & $K\cdot T/I$ & $K\cdot T/I$\tabularnewline
\hline 
weight change computations $C$ & $K\cdot T$ & $K\cdot T$ & $T$\tabularnewline
\hline 
history entry manipulations $H$ & $T$ & $T$ & $K\cdot T/I\cdot i$\tabularnewline
\hline 
\end{tabular}
\par\end{centering}
\caption{\textbf{Comparison of synapse update schemes.} From the view point
of a postsynaptic neuron, the table shows the maximal length of the
history $L$, the number of function calls $M$ of synapse code, the
number of computations $C$ of infinitesimal weight changes $\Delta W_{ij}(t^{\alpha},t^{\alpha+1})$,
and the number of history entry manipulations $H$ for a simulation
of $T$ time steps, a uniform inter-spike interval $I$ between spikes
of a single presynaptic neuron, and an in-degree $K$ for each neuron
and no delays. For the event-driven compression scheme the entries
show the length of the compressed history where $i$ is the number
of different spike times within the last inter-spike interval $I$.\label{tab:comparison}}
\end{table}

\begin{figure}
\begin{centering}
  \includegraphics[width=0.6\textwidth]{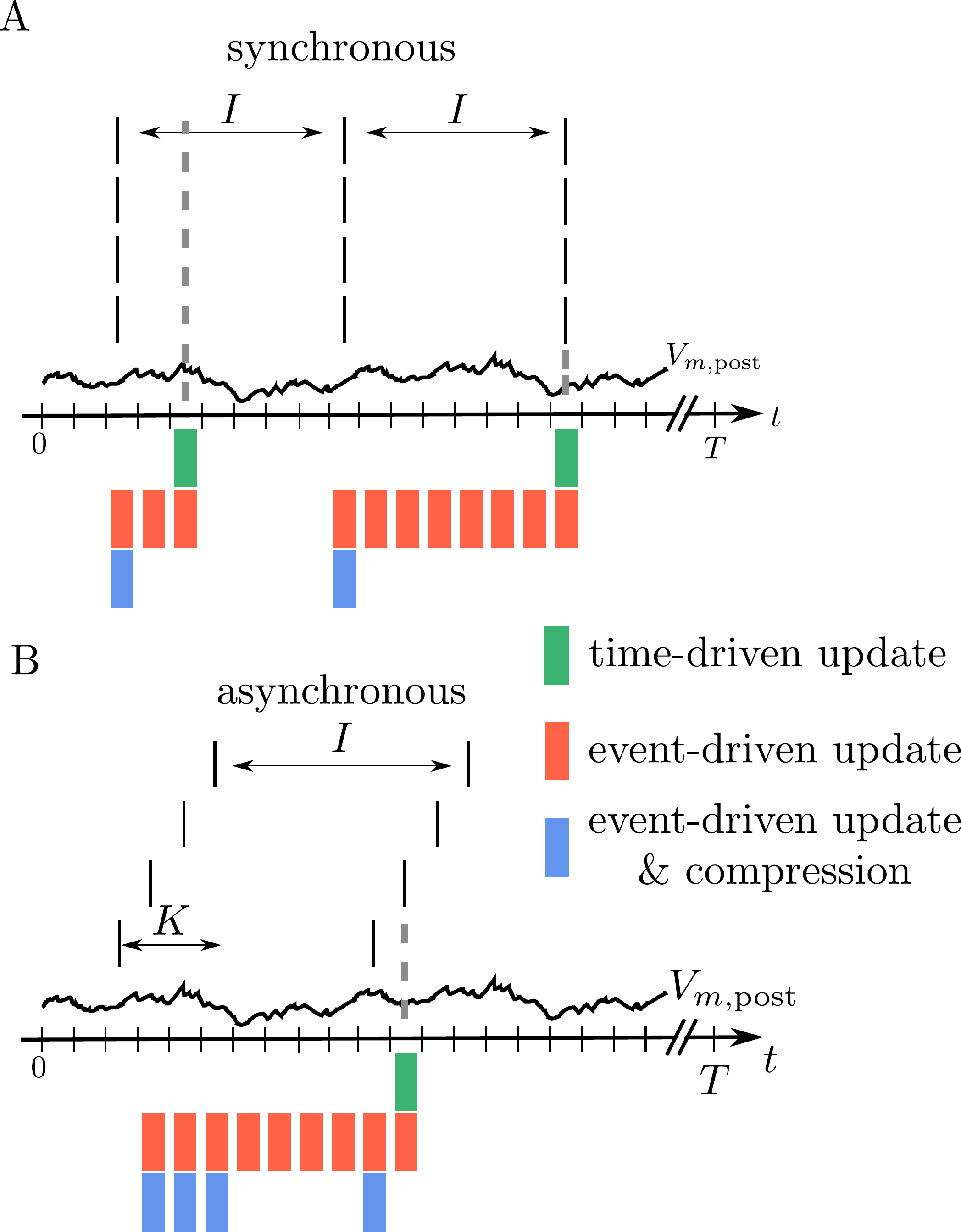}
\par\end{centering}
\caption{\textbf{Illustration of buffer sizes for different simulation schemes
in case of fully synchronous or asynchronous spikes.} \textbf{A} All
incoming spikes arrive synchronously: In the time-driven scheme the
synaptic weight is updated in every time step of the simulation, so
that only the current value of $V_{m,\mathrm{post}}$ needs to be
available (green). In the event-driven scheme every synapse processes
$V_{m,\mathrm{post}}$ from the last spike to the current one. Therefore,
the relevant time trace needs to be stored (yellow). In the compressed
event-driven scheme this part of $V_{m,\mathrm{post}}$ is processed
only once and used to update the weight of all the synapses. Since
the weight change is a function of the last spike time which is the
same for all the synapses, only one value needs to be updated (red).
In this situation the length $L$ of the compressed history is $i=1$,
see \prettyref{tab:comparison}. \textbf{B }All incoming spikes arrive
in different time bins: For the time-driven and the event-driven scheme
the scenario is similar to panel A. For the compressed event-driven
scheme the number of values that need to be updated equals the number
of incoming synapses $K$ so that $i=K$. \label{fig:spike_rasters}}
\end{figure}

Let's assume in the following that $t_{\mathrm{LS}}$ and $t_{\mathrm{S}}$
denote two consecutive spike times of a presynaptic neuron $j$. The
synaptic weight $W_{ij}(t_{\mathrm{S}})$ corresponding to the spike
at time $t_{\mathrm{S}}$ can be obtained from the weight $W_{ij}(t_{\mathrm{LS}})$
at the time of the previous spike at $t_{\mathrm{LS}}$ and \prettyref{eq:int_plasticity-1}
by employing \prettyref{eq:core} to calculate the latter. As $F$
mixes information of the pre- and postsynaptic neurons, this computation
should be done in the synapse. Since there are no spikes in between
$t_{\mathrm{LS}}$ and $t_{\mathrm{S}}$, it does not matter when
the synapse is performing the updates of its weight. Two possibilities
are: 1) Neurons calculate their own $s^{*}$ and $V^{*}$ for the
current time step and make it accessible to the synapse to enable
direct readout and update according to \prettyref{eq:core} in every
time step. This method corresponds to a time-driven update of synapses
(\prettyref{fig:Simulation-concepts-1}A). 2) Neurons store a history
of $s^{*}$ and $V^{*}$ and the synapse reads out this information
at $t_{\mathrm{S}}$, i.e. at the time where the weight update becomes
relevant for the network. This method corresponds to an event-driven
update of synapses (\prettyref{fig:Simulation-concepts-1}B). Both
methods have their advantages and disadvantages analyzed in the following
section. 

\subsubsection{Time-driven scheme\label{subsec:Time-driven-scheme}}

In a time-driven update scheme the information on the membrane potential
is directly processed by the synapses such that only the current value
of the membrane potential needs to be stored, corresponding to a membrane
potential history of length $L=1$ (\prettyref{fig:spike_rasters}
and \prettyref{tab:comparison}).  For a simulation of $T$ time
steps, the history needs to be manipulated $H=T$ times: the single
stored value gets updated once per time step. The price that comes
with the short history is that synapses need to be updated as often
as neurons. This amounts to $M=K\cdot T$ function calls to synapse
code for each neuron. Here $K$ denotes the in-degree of each neuron.
Each function call of synapse code causes a single computation of
$\Delta W_{ij}(t^{\alpha},t^{\alpha+1})$, giving rise to in total
$C=K\cdot T$ computations per neuron. The membrane potential trace
is thus effectively integrated $K$ times; once for each synapse.
As both $K$ and $T$ are large numbers in typical simulations of
plastic cortical networks, the amount of function calls and computations
is therefore large in this setting. The time-driven scheme furthermore
forces the execution of synapse code also at time steps where no update
would be required, i.e. at time steps, where $s_{i}^{*},s_{j}^{*},V_{i}^{*}$
have values for which $\Delta W_{ij}(t^{\alpha},t^{\alpha+1})=0$.
In addition, for delayed connections a history of $V_{i}^{*}$ of
length $L=d_{\max}$ of the maximal delay $d_{\max}$ measured in
simulation steps needs to be stored. We here assume the delay to be
on the postsynaptic side; it represents the time the fluctuations
of the somatic membrane potential propagate back through the dendrites
to the synapses. Therefore, $F$ does not depend on $V_{i}^{*}(t)$,
but on $V_{i}^{*}(t-d_{j})$ with a delay $d_{j}$ encoding the location
of the synapse with presynaptic neuron $j$.

\subsubsection{Event-driven scheme\label{subsec:Event-driven-scheme}}

In an event-driven update scheme for synapses, the time trace of
the membrane potential $V_{i}^{*}$ needs to be stored until all incoming
synapses have read out the information to update their weight for
a given period. The storage and management of such a history can be
expensive in terms of memory and runtime. In each time step, the
value of the current membrane potential is appended to the history,
leading to $H=T$ history manipulations for a simulation of $T$ time
steps. Assuming for simplicity a homogeneous inter-spike interval
of $I$ time steps between consecutive spikes of single neurons, we
in the following showcase some qualitative history sizes. As synapses
need all values of $V_{i}^{*}$ in between two consecutive spikes,
the maximum history length is $L=I$ (\prettyref{fig:spike_rasters}).
In case of different firing rates, $I$ corresponds to the maximum
inter-spike interval of any of the presynaptic neurons. Synapse code
in this scheme is, however, only called in the event of a spike, leading
to only $M=K\cdot T/I$ function calls per neuron, where $T/I$ is
the number of spikes passing a single synapse during the simulation
of $T$ time steps. The total amount of computations $C$ of weight
changes $\Delta W_{ij}(t^{\alpha},t^{\alpha+1})$ is of course unchanged
with respect to the time-driven scheme; they are just split across
less function calls ($C=M\cdot L=K\cdot T$). \prettyref{tab:comparison}
immediately shows the trade-off between memory consumption (length
of history) and run time (number of function calls): the event-based
scheme consumes more memory, but is faster than the time-driven scheme.
Note that since a history of the membrane potential is stored anyway,
this scheme is naturally applicable to connections with different
delays. A further performance increase can be achieved in plasticity
rules, where weight changes only happen under certain conditions on
$V_{i}^{*}$: if values $\Delta W_{ij}(t^{\alpha},t^{\alpha+1})\neq0$
are rare, a non-continuous history can be stored. In such a scenario,
time stamps need to be stored alongside the membrane potential to
enable synapses to read out the correct time intervals (see \prettyref{subsec:details_cl}).

\subsubsection{Event-driven compression\label{subsec:Data-compression}}

The event-driven compression scheme is a modified event-driven scheme
that makes use of the fact that for a specific class of plasticity
rules the integrated time trace of the membrane potential $V_{i}^{*}$
can be used to advance the weight update of all incoming synapses,
see \prettyref{subsec:Compression-methods}. Therefore, the time trace
of $V_{i}^{*}$ stored in the postsynaptic neuron only needs to extend
back to the last incoming spike (denoted by $t_{\mathrm{LI}}$ in
\prettyref{fig:Simulation-concepts} C). This way the history of $V_{i}^{*}$
is always short, as the total rate of incoming spikes is high in physiological
network states. Due to the dependence of the weight update on the
time of the last spike that crossed the synapse, the postsynaptic
neuron stores the compressed history of length $L=i$, where $i$
is the number of different spike times within the last inter-spike
interval $I$. (\prettyref{fig:spike_rasters}). The compressed history
is consequently never larger than the history length $L=I$ of the
ordinary event-driven scheme (\prettyref{fig:spike_rasters} B). For
synchronous spikes where the last presynaptic spike time is the same
for all synapses, the compressed history, however, contains only one
entry (\prettyref{fig:spike_rasters} A). Still, synapse code is executed
at every spike event, giving rise to $M=K\cdot T/I$ function calls.
The full membrane potential trace of length $T$ is effectively only
integrated once, amounting to in total $C=T$ infinitesimal weight
change computations that are performed in batches in between any two
incoming spike events (\prettyref{tab:comparison}). The price for
this is that history updates are more expensive: instead of appending
a single entry in each time step, at each spike event the full compressed
history is updated, giving rise to in total $H=M\cdot i=K\cdot T\cdot i/I$
history entry manipulations, as opposed to $H=T$ in the time- and
ordinary event-driven schemes \prettyref{tab:comparison}. In practice,
infinitesimal weight change computations are, however, often more
costly than history updates, such that the compression algorithm achieves
a performance increase  (see \prettyref{subsec:Performance}).

Finally, a drawback of the event-driven compression is that it relies
on the fact that all synapses use the same processed membrane potential
$V_{i}^{*}$. For distributed delays, $\Delta W_{i}(t_{LS},T)$ has
a dependence on the presynaptic neuron $j$ via $V_{i}^{*}(t-d_{j})$.
In this case, a separate compressed history needs to be stored for
every different delay of connections to the neuron.

\subsection{Reference implementation in network simulator with event-based synapse
updates\label{subsec:Reference-implementation}}

\noindent This section describes the implementation of two example
voltage-based plasticity rules by \citet{Clopath10_344} and \citet{Urbanczik14}
in a spiking neural network simulator that employs a time-driven update
of neurons and an event-based update of synapses. While the naming
conventions refer to our reference implementation in the simulation
software NEST, the algorithms and concepts presented below are portable
to other parallel spiking network simulators.

\begin{sloppypar}The Clopath and Urbanczik-Senn rule are chosen as
widely used prototypical models of voltage-based plasticity. The differences
in the two rules help to exemplify the advantages and disadvantages
of the algorithms discussed in \prettyref{subsec:Simulation-concepts}.
As originally proposed, they are implemented here for two different
types of neuron models, Ad-ex/Hodgkin-Huxley point-neurons for the
Clopath rule (\texttt{aeif\_psc\_delta\_clopath}/\texttt{hh\_psc\_alpha\_clopath})
and two-compartment Poisson neurons (\texttt{pp\_cond\_exp\_mc\_urbanczik})
for the Urbanczik-Senn rule. Extensions to multiple dendritic compartments
in the latter case are straight forward. Our implementation of \texttt{aeif\_psc\_delta\_clopath}
follows the reference implementation on ModelDB which introduced a
clamping of the membrane potential after crossing the spiking threshold
to mimic an action potential. Details can be found in \prettyref{subsec:Voltage-clamping}.

The plasticity rules differ in the state variable that is being stored
and its interpretation. For the Clopath rule, the stored variable
is a thresholded and filtered version of the membrane potential that
takes into account characteristics of membrane potential evolution
within cells in the vicinity of spike events. The restriction to
temporal periods around spikes suggests to implement a history that
is non-continuous in time. In contrast, the Urbanczik-Senn rule uses
the dendritic membrane potential to predict the somatic spiking; the
resulting difference is taken as an error signal that drives learning.
This error signal never vanishes and thus needs to be stored in a
time-continuous history.\end{sloppypar}

Finally, the proposed infrastructure for storing both continuous and
non-continuous histories is generic so that it can also be used and
extended to store other types of signals such as external teacher
signals.

\subsubsection{Exchange of information between neurons and synapses\label{subsec:Exchange-of-information}}

The implementation of voltage-based plasticity rules in NEST follows
the modular structure of NEST, key part of which is the separation
between neuron and synapse models. This separation makes it easy for
a newly added neuron model to be compatible with existing synapse
models and vice versa. A downside is that information, such as values
of parameters and state variables, is encapsulated within the respective
objects. Simulations in NEST employ a hybrid parallelization scheme:
OpenMP threads are used for intra node parallelization and the Message
Passing Interface (MPI) for inter node communication. In parallel
simulations, synapses are located at the same MPI process as the postsynaptic
neurons \citep{Morrison05a}. Thereby, no communication between MPI
processes is needed for the required transfer of information between
postsynaptic neurons and synapses to compute weight changes of connections
and only one spike needs to be communicated by a given source neuron
for all target neurons living on the same MPI process.

The model of STDP requires synapses to access spike times of postsynaptic
neurons. In order to provide a standardized transfer of this information
across all neuron models that support STDP, in recent years the so-called
\texttt{Archiving\_Node }has been introduced as a parent class of
the respective neuron models \citep{Morrison07_1437}. It provides
member functions to store and access spike histories. If a neuron
model supports STDP, it only needs to be a child of \texttt{Archiving\_Node
}and contain one additional line of code, namely a call of the function
\texttt{set\_spiketime()}, which stores the time of outgoing spike
events. We here extended this framework for voltage-based plasticity
rules and enhanced the functionality of the archiving node by the
member functions \texttt{write\_history(),} \texttt{get\_history()}
and \texttt{compress\_history()} to additionally store, read out and
manipulate voltage traces or other continuous signals (for Details,
see \prettyref{subsec:History-management}). To avoid overhead for
simulations with only STDP synapses, we introduced two child classes
of \texttt{Archiving\_Node}, \texttt{Clopath\_Archiving\_Node} and
\texttt{Urbanczik\_Archiving\_Node}, that each provide containers
and functions for the specific histories required for the two plasticity
rules. Neuron models that support the respective synapse model then
derive from the child classes instead of the root level archiving
node.

\begin{figure}
\begin{centering}
  \includegraphics[width=1\linewidth]{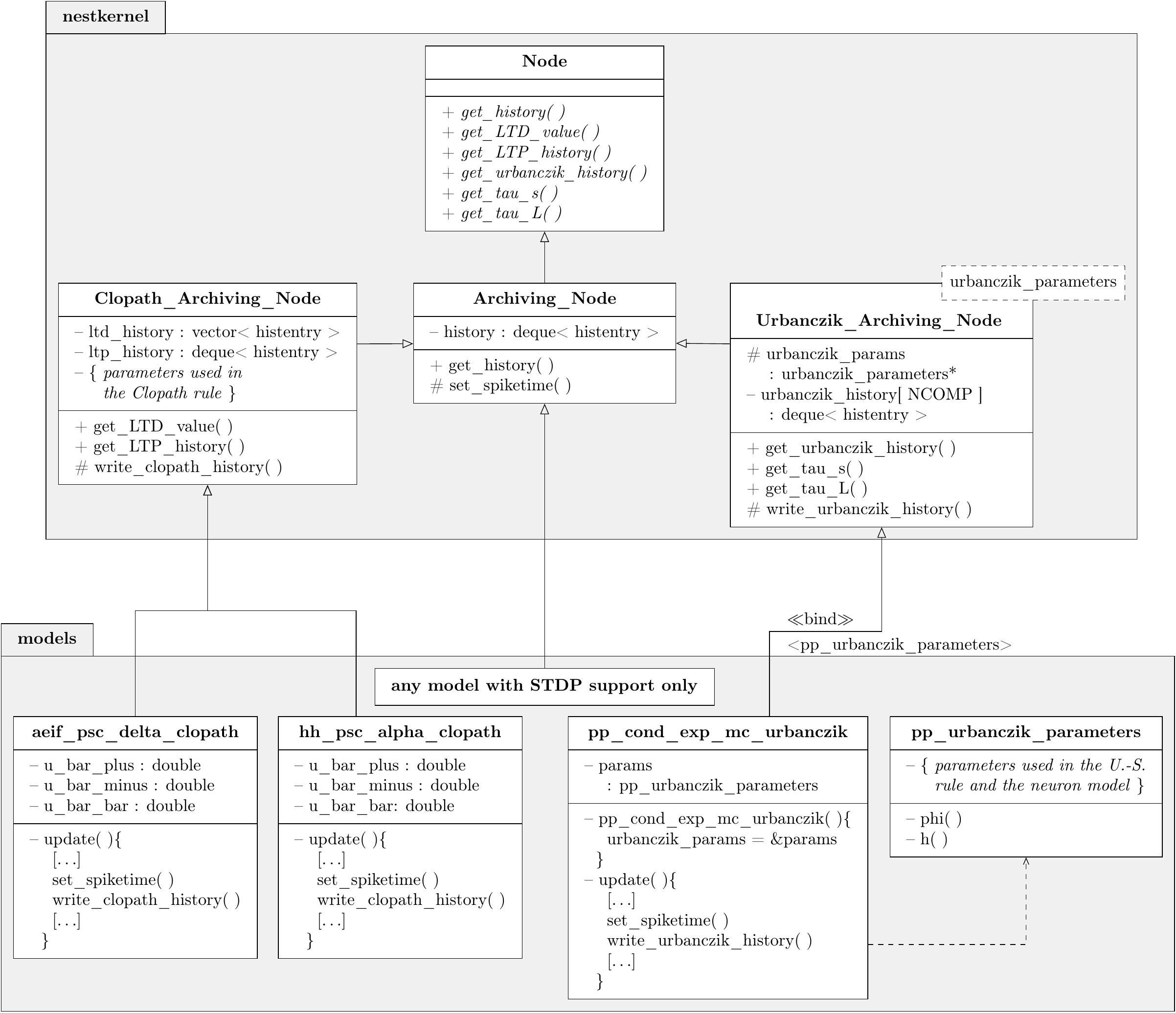}
\par\end{centering}
\caption{\textbf{Class diagram of NEST classes and functions.} Simplified class
diagram for embedding the Clopath (left) and Urbanczik-Senn rule (right)
in the NEST infrastructure. The code is distributed across the \texttt{nestkernel}
and neuron models. \texttt{nestkernel} contains the base class \texttt{Node}
of all neurons models. Models that support ordinary STDP are derived
from the \texttt{Archiving\_Node}, models that can use the Clopath
synapse (\texttt{aeif\_psc\_delta\_clopath} and \texttt{hh\_psc\_alpha\_clopath})
or Urbanczik-Senn synapse (\texttt{pp\_cond\_exp\_mc\_urbanczik})
are derived from the \texttt{Clopath\_Archiving\_Node} or the \texttt{Urbanczik\_Archiving\_Node},
respectively. The latter add the required functions for storing and
managing the history of continuous quantities. The model \texttt{pp\_cond\_exp\_mc\_urbanczik}
requires a helping class \texttt{pp\_urbanczik\_parameters} because
the \texttt{Urbanczik\_Archiving\_Node} needs to access functions
and parameters that are specific to the neuron model and therefore
not located in the \texttt{Urbanczik\_Archiving\_Node} to keep its
implementation more general.\label{fig:Simplified-class-diagramm}}
\end{figure}

\subsubsection{Delays and min\_delay communication\label{subsec:Delays-and-min_delay}}

All synapses implemented in NEST are so far purely event-driven. To
assess the performance of the time-driven update scheme of synapses
with voltage-based plasticity, we also implemented a time-driven version
of the Clopath and Urbanczik-Senn synapse. Spiking network simulators
exploit the delay of connections to reduce communication between compute
processes \citep{Morrison05a}: Instead of sending each spike individually,
spikes are buffered and sent in a batch after a certain period. The
length of this period, the \texttt{min\_delay}, corresponds to the
minimal delay of all connections in the network. The buffering of
spikes within this period is possible because the earliest time point
that a spike at time $t_{\mathrm{S}}$ can affect the postsynaptic
membrane potential is at $t=t_{\mathrm{S}}+\mathrm{min\_delay}$.
In between $t_{\mathrm{S}}$ and $t$ neurons are decoupled such that
their state variables can be propagated forward in time independent
of each other and in a batch \citep{Morrison08_267}. We implemented
the same \texttt{min\_delay} update scheme for synapses, by imposing
a function call to time-driven synapses in every \texttt{min\_delay}
period to update their synaptic weight. If \texttt{min\_delay} equals
the simulation step size $h$, this scheme corresponds to the scheme
explained in \prettyref{subsec:Time-driven-scheme}. Making use of
the \texttt{min\_delay} infrastructure of NEST speeds up simulations
with time-driven synapses in the case $d>h$ as fewer function calls
to synapses are needed (see \prettyref{subsec:Performance}). In case
of simulations with synaptic delays, the time-driven update scheme
requires the storage of a history of the membrane potential of length
\texttt{max\_delay}.

Storing state variables in event-driven schemes is more complex as
the history does not have a fixed length \texttt{max\_delay}. Instead
it needs to be dynamically extended and shortened. A long history
can occupy a large amount of memory and its processing by the synapses
becomes computationally expensive. Therefore, it is advantageous to
optimize the way how information is stored and accessed and how entries
that are no longer needed can be identified for deletion. For details
of these optimizations in our NEST implementation, see \prettyref{subsec:History-management}.

As discussed in \prettyref{subsec:Data-compression}, the event-based
compression scheme relies on the fact that all synapses to one postsynaptic
neuron employ the same $V_{i}^{*}$. This is not the case if delays
of the corresponding connections are distributed. The compression
scheme can therefore only be efficient if all delays have a fixed
value. If spikes are processed and synapses are updated in a chronological
order, then a well-defined segment of the history of $V_{i}^{*}$
can be integrated and the compressed history can be updated. In NEST,
spikes are, however, buffered within a period of \texttt{min\_delay}
before being sent and processed. Consequently, synapses are not necessarily
updated in chronological order. Therefore, the event-based compression
scheme can only be implemented in NEST in the case where delays equal
the simulation time step. Future work may explore whether the latter
restriction could be overcome by sorting all incoming spike events
of a given postsynaptic neuron prior to delivery.

\subsubsection{Specifics of Clopath plasticity\label{subsec:details_cl}}

We implement both an adaptive exponential integrate-and-fire neuron
model (\texttt{aeif\_psc\_delta\_clopath}) and a Hodgkin-Huxley neuron
model (\texttt{hh\_psc\_alpha\_clopath}) supporting Clopath plasticity.
These implementations consider the filtered versions $\bar{u}_{\pm}$
of the membrane potential as additional state variables of the neuron.
Thereby, they can be included in the differential equation solver
of the neurons to compute their temporal evolution. Parameters of
$\kappa_{\pm}$ consequently need to be parameters of the neuron object
rather than the synapse. The same is true for the values of $\theta_{\pm}$;
they are used in the neuron to determine whether $V_{i,\mathrm{LTP}}^{*}$
and $V_{i,\mathrm{LTD}}^{*}$ evaluate to zero, which systematically
happens due to the Heaviside functions in their definitions.

The LTD mechanism is convenient to implement within the event-driven
framework: when the synapse is updated at time $t$, it reads the
values $\bar{u}_{-}\left(t-d\right)$ and $\theta_{-}$ from its target
and computes the new weight. Here, $d$ denotes the dendritic delay
of the connection that accounts for the time it takes to propagate
somatic membrane potential fluctuations to the synapse. The archiving
node contains a cyclic buffer, also called ring buffer, that stores
the history of $\bar{u}_{-}$ for the past \texttt{max\_delay} time
steps so that the synapse can access a past value of this quantity.
Consequently, the LTD history is always short and can be forgotten
in a deterministic fashion.

The computation of the weight change due to LTP requires the evaluation
of the integral over $V_{i,\mathrm{LTP}}^{*}(t)$. The latter is stored
in the archiving node as a vector whose elements are objects that
contain three values: the corresponding time $t$, the value of $V_{i,\mathrm{LTP}}^{*}$
and an access counter that initially is set to zero.

\paragraph{Time-driven update:}

For simulations with homogeneous delays equal to the simulation time
step, the history of $V_{i,\mathrm{LTP}}^{*}$ always contains only
a single value as it is read out in every time step by all synapses.
For larger delays, the history is of length \texttt{max\_delay}, and
each synapse reads out a segment of length \texttt{min\_delay}, increasing
the access counter of the corresponding entries by one. For the last
synapse that requests a certain segment, the access counter then equals
the in-degree $K$, which is the criterion to delete the corresponding
entries from the history. Although for simplicity done in our reference
implementation, the time-driven scheme does not require us to store
the time stamp $t$ of each history entry. The overhead of this additional
number is, however, negligible.

\paragraph{Event-driven update:\label{par:clopath_specifics_ed}}

In event-driven schemes, the history of $V_{i,\mathrm{LTP}}^{*}$
dynamically grows and shrinks depending on the spikes of presynaptic
neurons. Since many values of $V_{i,\mathrm{LTP}}^{*}$ are zero,
it is beneficial to only store the non-zero values. In this case,
a time stamp of each entry is required to assign values of the non-continuous
history of $V_{i,\mathrm{LTP}}^{*}$ to their correct times. In case
of the uncompressed scheme, when a synapse $j$ is updated at time
$t_{S}$ of a spike, it requests the part of the history between the
last spike $t_{LS}$ and the current spike $t_{S}$ (minus the dendritic
delay, see \prettyref{subsec:History-management}) from the archiving
node. This history segment is then integrated in synapse $j$ and
used for its weight update. Each synapse thus integrates the history
$V_{i,\mathrm{LTP}}^{*}$ anew (\prettyref{subsec:Event-driven-scheme}).
For the compressed scheme, the history of $V_{i,\mathrm{LTP}}^{*}$
is integrated between the last incoming spike at $t_{\mathrm{LI}}$
and the current spike at $t_{\mathrm{S}}$ inside the archiving node.
Using this newly integrated time trace, the weight of synapse $j$
is updated and the compressed history for all other last spike times
is advanced. Afterwards the history of $V_{i,\mathrm{LTP}}^{*}$ is
deleted. Thereby, $V_{i,\mathrm{LTP}}^{*}$ is only integrated once
for all synapses.

In any case, the integrated history of $V_{i,\mathrm{LTP}}^{*}$ needs
to be combined with the presynaptic spike trace $s_{j}^{*}$. The
latter is easily computed analytically inside the synapse because
it is an exponential decay of the corresponding value at the time
of the last spike. At the end of the update process the trace is increased
by $\tau_{s}^{-1}$ to account for the trace of the current spike,
where $\tau_{s}$ is the time constant of the kernel $\kappa_{s}$.

\subsubsection{Specifics of Urbanczik-Senn plasticity}

Following the original publication \citep{Urbanczik14}, we implement
a Poisson spiking neuron model (\texttt{pp\_cond\_exp\_mc\_urbanczik})
supporting Urbanczik-Senn plasticity. One peculiarity of this model
is that the gain function $\phi$ that translates the membrane potential
into a firing rate also enters the plasticity rule through $V^{*}$.
Therefore $\phi$ as well as its parameters need to be known by the
neuron and the synapse. Creating an additional helper class (\texttt{pp\_urbanczik\_parameters})
as a template argument for the corresponding archiving node (\texttt{Urbanczik\_Archiving\_Node})
and neuron model (\texttt{pp\_cond\_exp\_mc\_urbanczik}) solves this
problem (\prettyref{fig:Simplified-class-diagramm}): it contains
all parameters and functions required by both classes.As explained
in \prettyref{subsec:Example_US}, the representation \prettyref{eq:us_weight_change}
is more beneficial for implementing the Urbanczik-Senn rule than that
of \prettyref{eq:us_weight_bad}. The first two integrals in \prettyref{eq:us_weight_change}
only extend from $t$ to $T$; history entries for times smaller than
$t$ are not needed and can be deleted after the corresponding update.
The dependence on the full history back until $0$ arising from the
convolution with $\kappa$ is accumulated in the last term in \prettyref{eq:us_weight_change},
which the synapse computes with the help of storing one additional
value $I_{2}\left(0,t\right)$. At the end of a weight update this
value is overwritten by the new value $I_{2}\left(0,T\right)=e^{-\frac{T-t}{\tau_{\kappa}}}I_{2}\left(0,t\right)+I_{2}\left(t,T\right)$
which is then used in the next update.Either the synapse (time- and
event-driven update) or the archiving node (event-driven compression)
compute the two integrals $I_{1}$ and $I_{2}$ but in all cases the
archiving node stores the history of $V_{i}^{*}\left(t\right)$.

\subsection{Reproduction of results in literature\label{subsec:Reproduction-of-results}}

\begin{figure}
\begin{centering}
  \includegraphics[width=0.75\textwidth]{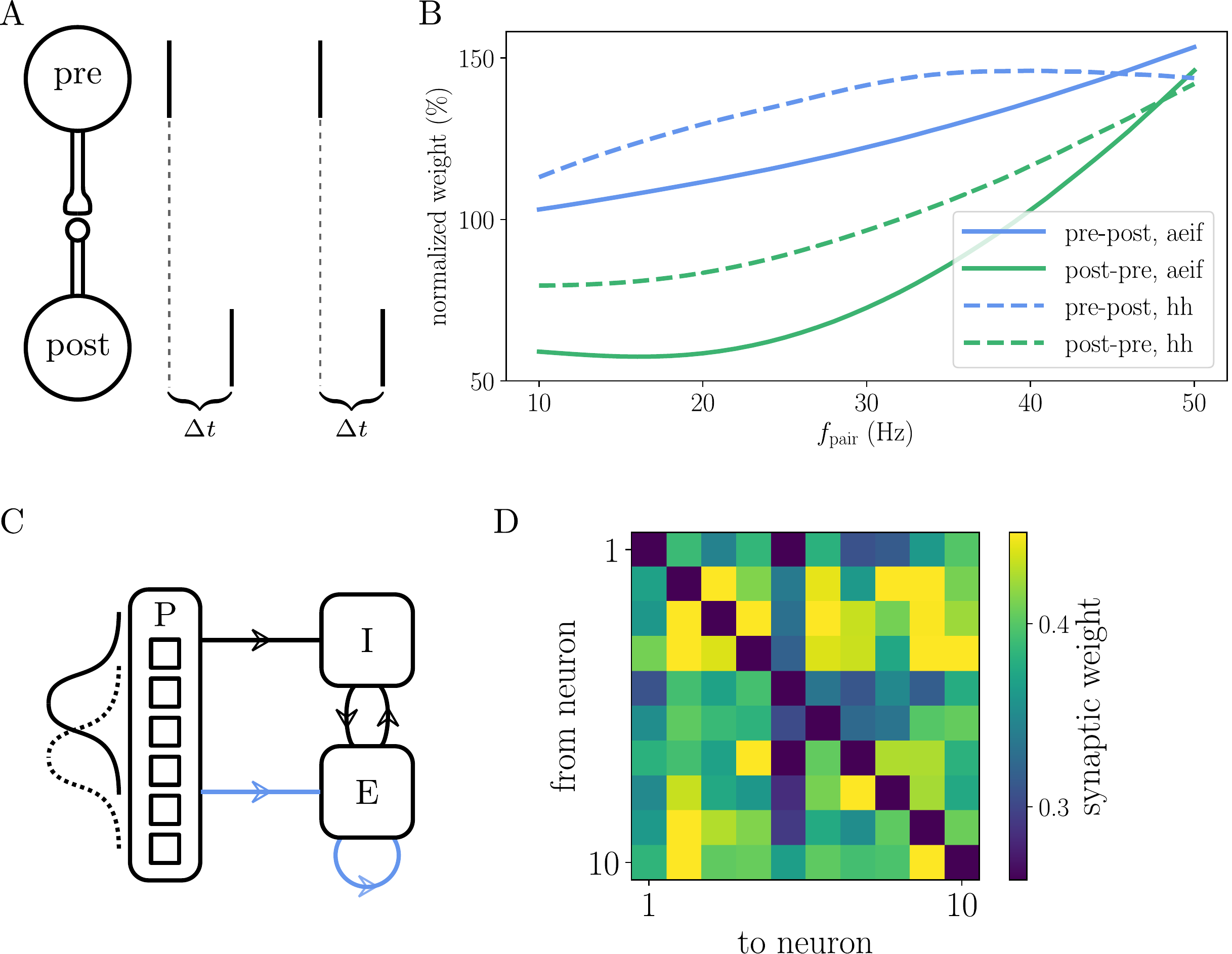}
\par\end{centering}
\caption{\textbf{Reproduction of results with Clopath rule. A} Setup of the
spike pairing experiment. Two neurons (``pre'' and ``post'') that
are connected by a plastic synapse receive input so that they spike
one after another with a delay $\Delta t$. The change of the synaptic
weight is computed according to the Clopath rule as a function of
the frequency $f_{\mathrm{pair}}$ with which the spike pairs are
induced. \textbf{B} Result of the spike pairing experiment. The relative
change of the synaptic weight after five spike pairs as a function
of $f_{\mathrm{pair}}$ is shown for two different neuron models (aeif:
solid lines, Hodgkin-Huxley: dashed lines). The blue lines represent
a setup where the postsynaptic neuron fires after the presynaptic
one (pre-post, $\Delta t=10\,\mathrm{ms}$) and the green lines represent
the opposite case (post-pre, $\Delta t=-10\,\mathrm{ms}$). This panel
corresponds to figure 2b of \citet{Clopath10_344}. \textbf{C} Setup
of the network that produces strong bidirectional couplings. The network
consists of an inhibitory (I) and an excitatory (E) population which
receive Poisson spike trains (P) as an external input. The firing
rate of the latter is modulated with a Gaussian shape whose center
is shifted every $100\,\mathrm{ms}$. The external input connections
to the excitatory population are plastic as well as the connections
within the excitatory population (indicated by blue arrows). \textbf{D}
Synaptic weights of the all-to-all connected excitatory neurons after
the simulation of the network. Strong bidirectional couplings can
be found, e.g. between neurons 2 and 3, 2 and 9, and 4 and 7. The
setup of this experiment is similar to that shown in figure 5 of \citet{Clopath10_344}.
A more detailed description of the two experiments can be found in
\prettyref{subsec:Implementation-of-experiments}. \label{fig:clopath_results}}
\end{figure}

The reference implementation of the Clopath plasticity reproduces
the results from \citet{Clopath10_344} on the frequency dependence
of weight changes in spike-pairing experiments and the emergence of
bidirectional connections in small all-to-all connected networks (\prettyref{fig:clopath_results}).
The setup of the spike-pairing experiment in \prettyref{fig:clopath_results}A
consists of two neurons connected via a plastic synapse. The pre-
and postsynaptic neuron are forced to spike with a time delay of $\Delta t$
multiple times which leads to a change in synaptic weight that depends
on the frequency of the spike pairs (\prettyref{fig:clopath_results}B).
The setup of the small network is shown in \prettyref{fig:clopath_results}C.
The weights of the plastic synapses within the recurrently connected
excitatory population are initialized all to the same value. At the
end of the simulation during which the network receives a time varying
input, some pairs of neurons show strong bidirectional connections
(\prettyref{fig:clopath_results}D). See Appendix \prettyref{subsec:Implementation-of-experiments}
for details on the setup of both experiments as implemented in NEST.

\begin{figure}
  \centering{}\includegraphics[width=1\textwidth]{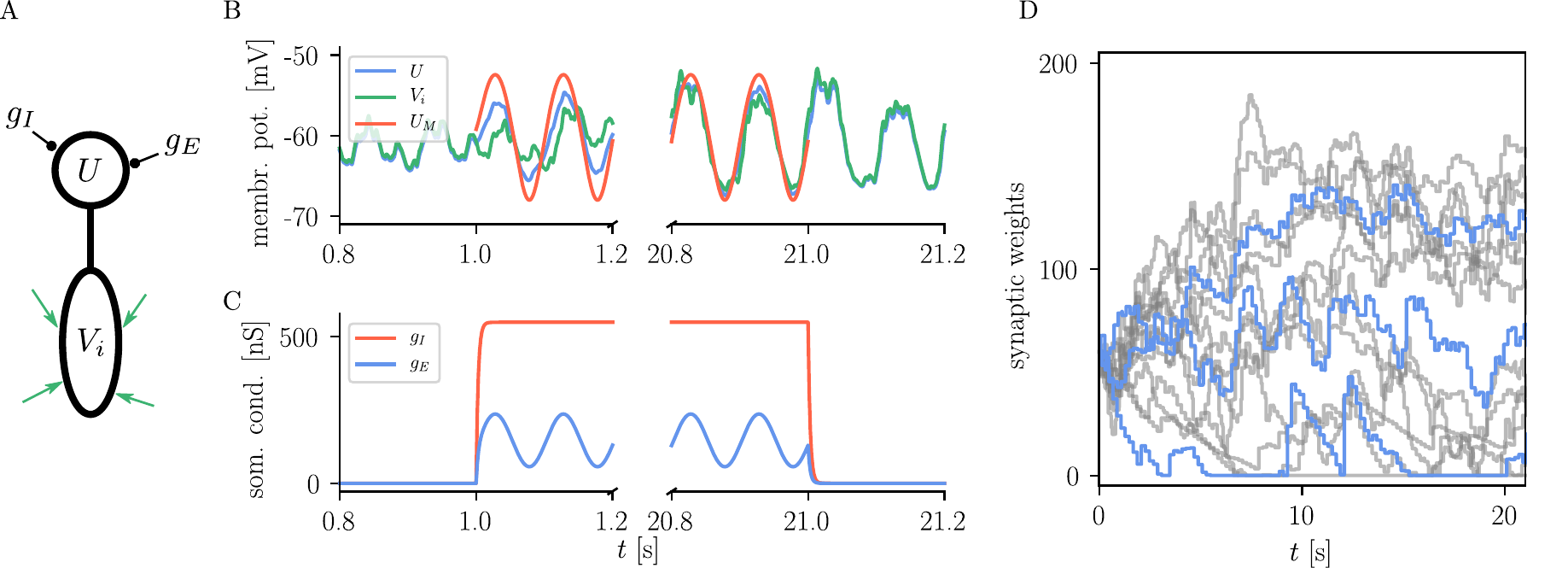}\caption{\textbf{Reproduction of results with Urbanczik-Senn rule. A} Setup
of a simple learning task using the Urbanczik-Senn plasticity rule.
The somatic conductances $g_{I}$ and $g_{E}$ of a two-compartment
neuron are modulated such that they induce a teaching signal with
sinusoidal shape. The dendrite receives a repeating spike pattern
as an input via plastic synapses (green arrows). \textbf{B} The synapses
adapt their weights so that the somatic membrane potential $U$ (blue)
and the dendritic prediction $V_{i}$ (green) follow the matching
potential $U_{M}$ (red) after learning. \textbf{C} Excitatory ($g_{E}$)
and inhibitory ($g_{I}$) somatic conductances that produce the teaching
signal. A and B corresponds to figure 1 of \citet{Urbanczik14}. \textbf{D}
Temporal evolution of the synaptic weights during learning. For the
sake of better overview, only a subset of weights is shown (gray)
with three randomly chosen time traces highlighted in blue. Synapses
in NEST fulfill Dale's principle which means that a weight update
cannot convert an excitatory into an inhibitory synapse and vice versa
giving rise to the rectification at zero. \label{fig:US-reproduction}}
\end{figure}

The basic use of the Urbanczik-Senn rule in NEST is exemplified in
\prettyref{fig:US-reproduction} which shows the reproduction of a
simple learning experiment from the original publication \citep{Urbanczik14}.
Here the neuron is supposed to transform spike patterns in the input
to the dendritic compartment into a sinusoidal modulation of the somatic
membrane potential. This target potential is determined by an external
teaching signal during learning. Via minimizing the error between
the dendritic prediction of the somatic membrane potential and the
actual somatic membrane potential, weights of dendritic synapses are
organizing such that the neuron can produce the desired membrane potential.
There is, however, no stop-learning region in the Urbanczik-Senn rule
(for a modified version, see \citealt{Cartiglia20_84}): The error
never vanishes completely which causes weights to keep changing despite
an overall good approximation of the target signal. Details of the
experiment and NEST setup can be found in Appendix \prettyref{subsec:Implementation--US}.

\subsection{Performance of the reference implementations\label{subsec:Performance}}

\subsubsection{Clopath plasticity\label{subsec:Clopath-performance}}

In order to evaluate the performance of the implementation of the
Clopath rule in NEST, in a weak-scaling setup, we simulate excitatory-inhibitory
networks of increasing size, but fixed in-degree $K$. As we expect
the performance to critically depend on the number of synapses, we
examine two scenarios: a small in-degree $K=100$, and a rather large
in-degree $K=5000$. While the first case might be suitable for small
functional networks, the latter in-degree represents a typical number
for cortical networks. Further details on network and simulation parameters
are given in supplementary Table S5. As a reference, we also simulate
the same network with STDP synapses, which require much less computations
as they rely solely on spike times. To achieve the same network state,
that is the same spikes, for the different connectivity rules, we
impose the weights to stay constant across time by setting learning
rates to zero. This way all computations for weight changes are being
performed, but just not applied. This has the additional advantage
that reasonable asynchronous irregular network states are simple to
find based on predictions for static synapses \citep{Brunel00}.

\begin{figure}
\begin{centering}
  \includegraphics[width=1\linewidth]{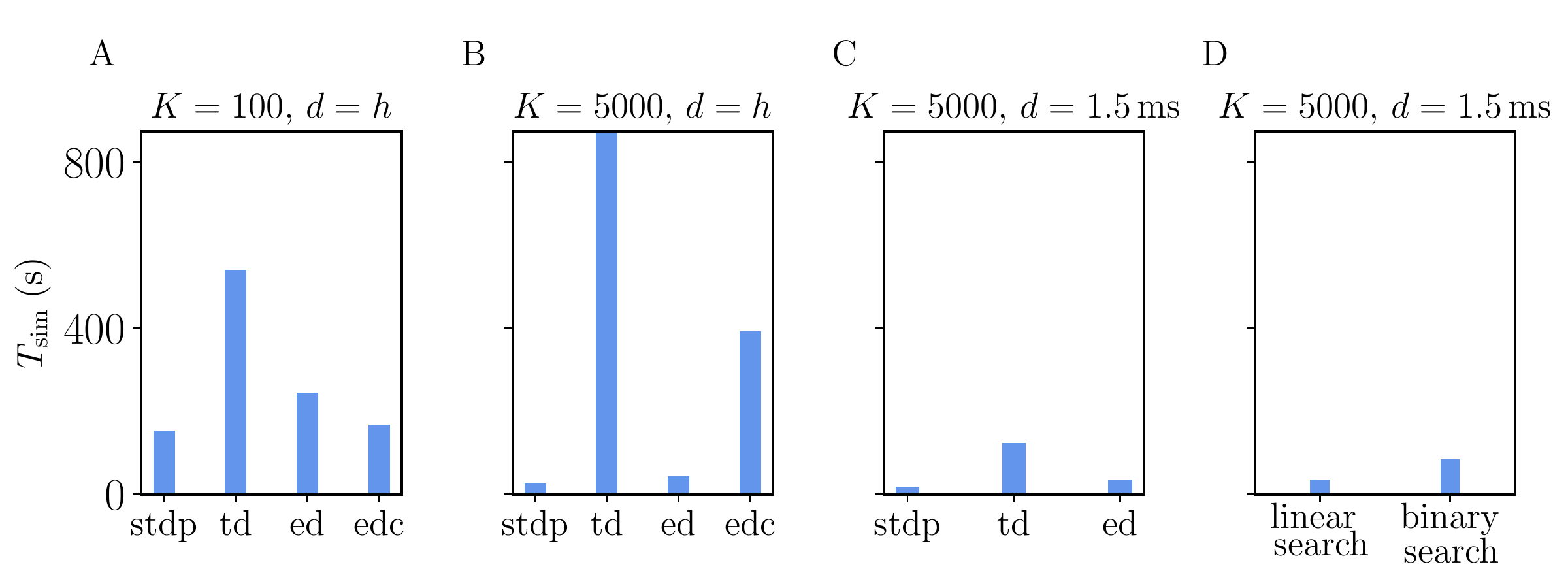}
\par\end{centering}
\caption{\textbf{Comparison of simulation times $T_{\mathrm{sim}}$ for excitatory-inhibitory
networks with different implementations of the Clopath plasticity
in NEST.} Simulation times exclude network building and only account
for updates of the dynamical state of the system. The following implementations
are shown: ``stdp'': standard implementation of STDP synapse, ``td'':
time-driven implementation of Clopath synapse, ``ed'': event-driven
scheme as included in NEST 2.18, ``edc'': event driven compression.
\textbf{A} Network of size $N=1.92\cdot10^{6}$ with small in-degree
$K=100$ and all synapses having a delay $d$ equal to the resolution
of the simulation $h=0.1\,\mathrm{ms}$. \textbf{B} Network of size
$N=1.54\cdot10^{5}$ with large in-degree $K=5000$ and $d=h$. \textbf{C}
Same network as in panel B but $d=1.5\,\mathrm{ms}$ (for $d>h$ ``edc''
not compatible with NEST, see \prettyref{subsec:Data-compression}).
In A, B, and C both ``ed'' and ``edc'' use linear search of the
history and access counters, see \ref{subsec:History-management}.
\textbf{D} Comparison between ``ed''-implementations using linear
search and binary search of the history. All simulations use $768$
threads distributed over $32$ compute nodes each running one MPI
process. Further parameters as in supplementary Table S5.\label{fig:clopath_performance}}
\end{figure}

The Clopath rule has originally been proposed for connections without
delays \citep{Clopath10_344}. Therefore, we first evaluate its performance
in this setting (delay equals simulation time step), which is, however,
not the natural setting for a simulator like NEST that makes use of
delays to speed up communication between compute processes. The first
observation is that, as expected, simulations with Clopath synapses
are slower than those with ordinary STDP (\prettyref{fig:clopath_performance}).
Given the update of synapses in every simulation step, the time-driven
scheme for Clopath synapses is much slower than the event-driven scheme
(\prettyref{fig:clopath_performance}A). The difference becomes larger
the more synapses there are (\prettyref{fig:clopath_performance}B).
Introducing a delay leads to less function calls to synapses (once
every \texttt{min\_delay}) and therefore increases the speed of the
time-driven scheme (\prettyref{fig:clopath_performance}C). Its simulation
times, however, remain much above the event-driven scheme. This comparison
illustrates the benefit of event-driven updates for Clopath synapses.

\begin{figure}
\begin{centering}
  \includegraphics[width=0.33\linewidth]{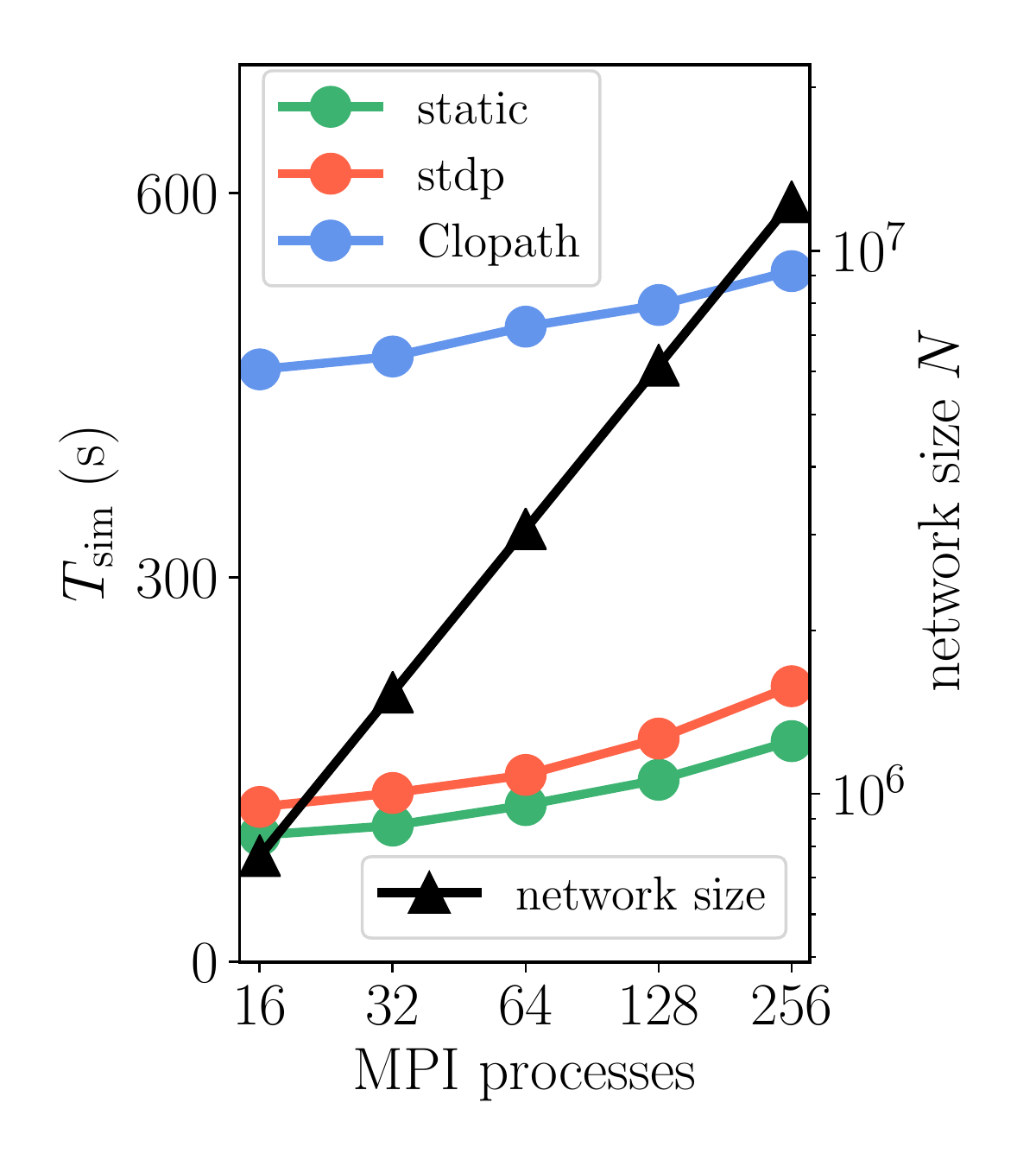}
\par\end{centering}
\caption{\textbf{Scaling of simulation time $T_{\mathrm{sim}}$ with network
size for $2\,s$ of biological time: Clopath plasticity.} Weak scaling:
computational resources (horizontal axis) increase proportionally
to network size $N$ (black curve and triangles, right vertical axis).
Event-driven scheme for Clopath rule (blue) compared to static (green)
and STDP synapse (red). Network and simulation parameters as in supplementary
Table S5 with in-degree $K=5000$. For all simulations each compute
node runs one MPI process with 24 threads. \label{fig:scaling}}
\end{figure}

How does compression of the history change the picture? As discussed
in \prettyref{subsec:Data-compression}, compression has the advantage
of not integrating the membrane potential history for each synapse
separately. A downside of the event-based compression is that it requires
storing one history entry for each last spike time of presynaptic
neurons. For large in-degrees, this history is therefore longer than
the history of $V_{i,\mathrm{LTP}}^{*}$, which we implemented as
non-continuous for the Clopath rule. Consequently, the event-based
compression scheme only outperforms the ordinary event-driven scheme
for small in-degrees (\prettyref{fig:clopath_performance}A), but
not for large in-degrees (\prettyref{fig:clopath_performance}B).
Given that the compression can only be implemented in NEST for connections
with delay equal to the resolution of the simulation (see \prettyref{subsec:Delays-and-min_delay}),
the method of choice is therefore the ordinary event-driven scheme
(\prettyref{subsec:Event-driven-scheme}). Although a bit slower,
its run-time is on the same order of magnitude as the ordinary STDP
synapse, with similar weak-scaling behavior (\prettyref{fig:scaling}).
The additional computations with respect to STDP result in a constant
overhead.

Another advantage of having short non-continuous histories is that
searching the history at readout is fast. A simple linear iteration
scheme is therefore even faster than a binary search (\prettyref{fig:clopath_performance}D)
because the latter search requires an additional list of presynaptic
spike times (see \prettyref{subsec:History-management}) which is
unnecessary overhead in this scenario. As a result the ordinary event-driven
scheme with linear history iteration is the most general and efficient
scheme and therefore integrated into NEST 2.18 \citep{Nest2180}.

\subsubsection{Urbanczik-Senn plasticity}

\begin{figure}
\begin{centering}
  \includegraphics[width=1\linewidth]{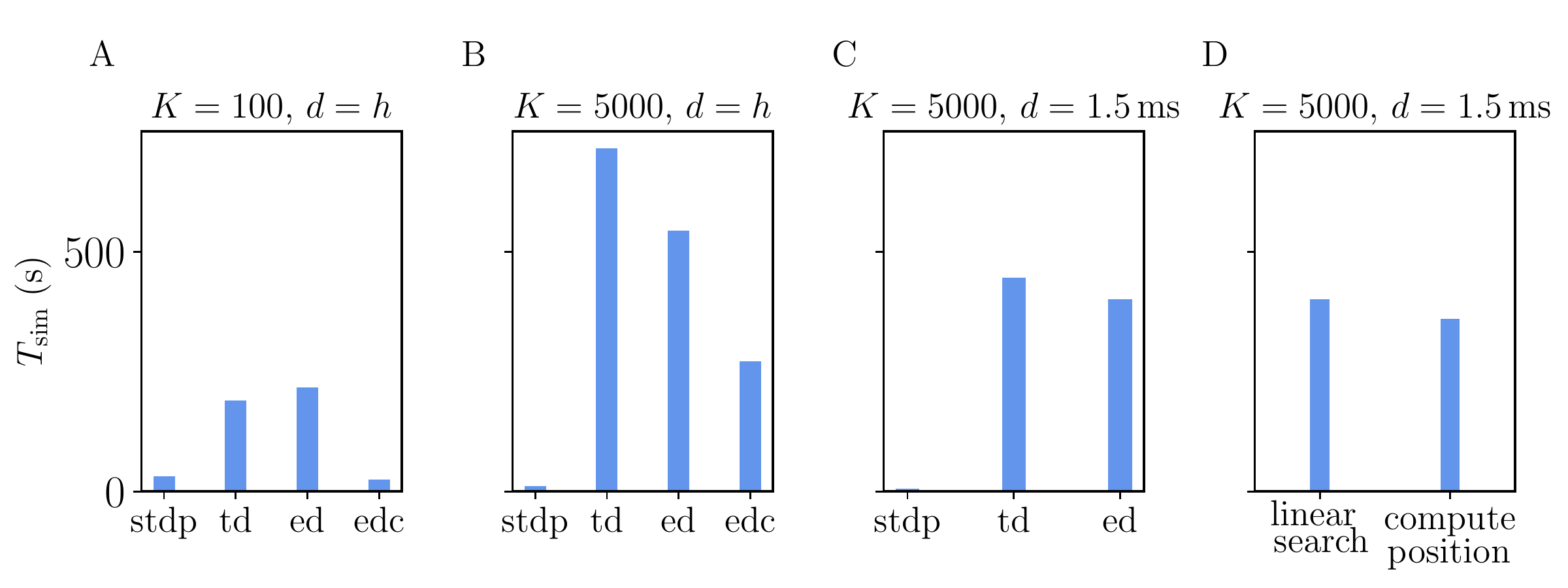}
\par\end{centering}
\caption{\textbf{Comparison of simulation times $T_{\mathrm{sim}}$ for excitatory-inhibitory
networks with different implementations of the Urbanczik-Senn plasticity
in NEST.} The following implementations are shown: ``stdp'': standard
implementation of STDP synapse in NEST, ``td'': time-driven implementation
of Urbanczik-Senn synapse, ``ed'': event-driven scheme, ``edc'':
event driven compression. \textbf{A} Network of size $N=3.84\cdot10^{5}$
with small in-degree $K=100$ and all synapses having a delay $d$
equal to the resolution of the simulation $h=0.1\,\mathrm{ms}$. \textbf{B}
Network of size $N=3.84\cdot10^{4}$ with large in-degree $K=5000$
and $d=h$. \textbf{C} Same network as in panel B but $d=1.5\,\mathrm{ms}$
(for $d>h$ ``edc'' not compatible with NEST, see \prettyref{subsec:Data-compression}).
In A, B, and C both ``ed'' and ``edc'' use linear search of the
history and the access counters, see \ref{subsec:History-management}.
\textbf{D} Comparison between ``ed''-implementations using linear
search and binary search of the history. All simulations use $768$
threads distributed over 32 compute nodes each running one MPI process.
Details on network parameters in Table S5 in supplement.\label{fig:US-performance}}
\end{figure}

The Urbanczik-Senn rule, in its original version, does not account
for delays in connections \citep{Urbanczik14}. As for the Clopath
rule, we therefore first evaluate its performance for connections
with delays that equal the simulation time step. We compare the results
to networks with ordinary STDP synapses, again setting all learning
rates to zero to maintain the same network state across different
types of plasticity. Naturally, the processing of the membrane potential
information makes the Urbanczik-Senn plasticity less efficient to
simulate than networks with ordinary STDP synapses (\prettyref{fig:US-performance}).
Note that the absolute numbers of simulation times are not directly
comparable to simulations with Clopath plasticity (\prettyref{fig:clopath_performance})
as network sizes are smaller here (Table S5 in supplement). Networks
with small and large in-degrees behave qualitatively similar: given
the long continuous history that needs to be stored and read out,
the event-driven scheme does not significantly outperform the time-driven
scheme (\prettyref{fig:US-performance}A,B). In the network with small
in-degree, the time-driven scheme is even slightly faster (\prettyref{fig:US-performance}A).
This behavior reverses for large in-degrees as the number of synapse
calls grows stronger than the length of the history (\prettyref{fig:US-performance}B).
However, given that the length of the history is so critical in this
rule, the compression algorithm can in both cases achieve a significant
increase in performance (\prettyref{fig:US-performance}A,B). This
performance increase is larger the smaller the in-degree, as the compressed
history becomes shorter (\prettyref{fig:US-performance}A). Due to
current NEST specifics (see \prettyref{subsec:Delays-and-min_delay}),
the compression algorithm cannot be used in settings with delays that
are are larger than the simulation time step (\prettyref{fig:US-performance}C):
Here, as expected, the time-driven scheme becomes faster than in the
$d=h$ case, but it is in general still comparable in performance
to the event-driven scheme. The latter is therefore the method of
choice for simulations with delayed connections; for zero-delay connections,
the compression algorithm performs best. Whether the history readout
is done via linear iteration or via computing positions of history
entries has no significant impact on the performance (\prettyref{fig:US-performance}D).
Therefore, the simple linear iteration is integrated in NEST 3.

We furthermore employ a weak-scaling setup with excitatory-inhibitory
networks of increasing size and fixed in-degree $K=5000$ (\prettyref{fig:scaling-1}A,B,
and Table S5 in supplement). Apart from a constant offset, the scaling
of simulation time $T_{\mathrm{sim}}$ for updating neurons and synapses
is similar for Urbanczik, static and STDP synapses. With increasing
network size $N$ and proportionally increasing number of MPI processes,
$T_{\mathrm{sim}}$ rises only slightly (\prettyref{fig:scaling-1}B),
indicating almost ideal weak-scaling behavior. The constant offset
in $T_{\mathrm{sim}}$ is larger than for Clopath synapses as the
Urbanczik-Senn rule requires longer histories of membrane potentials
and a more extensive history management.

\begin{figure}
\begin{centering}
  \includegraphics[width=1\linewidth]{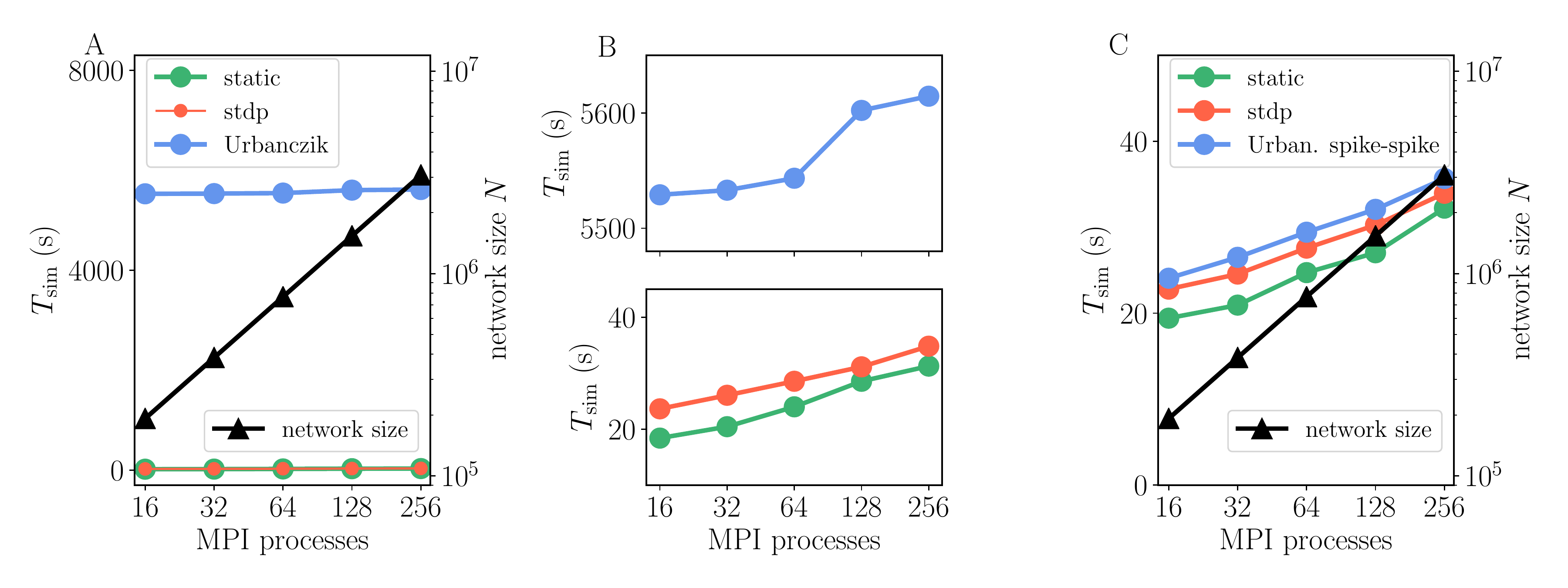}
\par\end{centering}
\caption{\textbf{Scaling of simulation times $T_{\mathrm{sim}}$ with network
size for $2\,s$ of biological time: Urbanczik-Senn plasticity.} Same
weak scaling as in \prettyref{fig:scaling}. \textbf{A} Event-driven
Urbanczik-Senn rule (blue) compared to static (green) and STDP synapse
(red). On the scale of the vertical axis the red curve (STDP synapses)
falls on top of the green curve (static synapses), indicated by finer
line width and marker size of the former . \textbf{B} Same simulation
time data as in A but with a smaller range on the vertical axis. Upper
panel: enlargement of Urbanczik-Senn data. Lower panel: enlargement
of data for static and STDP synapses. \textbf{C} Spike-spike version
of the Urbanczik-Senn rule compared to static and STDP synapse. Network
and simulation parameters as in supplementary Table S5 with in-degree
$K=5000$. For all simulations each compute node runs one MPI process
with 24 threads.\label{fig:scaling-1}}
\end{figure}

\subsection{Conclusions}

\label{subsec:Conclusions}

The analyses of the Clopath and the Urbanczik-Senn plasticity as prototypical
examples for rules that rely on storage of discontinuous versus continuous
histories show that the former are much faster to simulate, in particular
for large networks that require distributed computing. For discontinuous
histories, the event-driven scheme is most generally applicable and
efficient, which makes corresponding rules easy to integrate into
modern simulators with event-based synapses. The performance gap between
the different rules should be kept in mind in the design of new learning
rules. Furthermore, it is worthwhile to test modifications of existing
learning rules to decrease the amount of stored information.

For illustration, we here test a spike-based alternative to the original
Urbanczik-Senn rule, where we replace the rate prediction $\phi\left(V_{i}\left(t\right)\right)$
in $V^{*}$ of \prettyref{eq:prediction_error} by a noisy estimate,
which we generate by a non-homogeneous Poisson generator with rate
$\phi\left(V_{i}\left(t\right)\right)$, see \prettyref{subsec:US_experiment_modified}.
The prediction error then results in a comparison of somatic and dendritic
spikes, $s_{i}$ and $s_{i}^{\mathrm{dend}}$, respectively; it is
therefore purely based on point processes. In terms of storage and
computations, the rule thereby becomes similar to ordinary STDP (cf.
\prettyref{eq:int_STDP}). This becomes apparent in the weak-scaling
experiment in \prettyref{fig:scaling-1}C, which shows that the modification
of the learning rule results in a speedup of a factor $10$ to $30$
arriving essentially at the same run time as the ordinary STDP rule.

When changing learning rules to improve the efficiency of an implementation,
the question is in how far the modified rule, in our example including
the noisy estimate of the dendritic prediction, still fulfills the
functionality that the original rule was designed for. Generally,
without control of the error any simulation can be made arbitrarily
fast. Therefore \citet{Morrison07_47} define efficiency as the wall-clock
time required to achieve a given accuracy. We test in the appendix
(\prettyref{fig:US_modified}) whether the dynamics is still robust
enough to achieve proper learning and function in the reproduced task
of \prettyref{fig:US-reproduction}. The learning works as well as
in the original Urbanczik-Senn rule. However, given the simplicity
of the chosen task , this result may not generalize to other more
natural tasks. We leave a more detailed investigation of this issue
to future studies. The basic exploration here, however, illustrates
how taking into account the efficiency of implementations can guide
future development of learning rules to make them practically usable
for large-scale simulations of brain networks.

\section{Discussion}

\label{sec:Discussion}

This work presents efficient algorithms to implement voltage-based
plasticity in modern neural network simulators that rely on event-based
updates of synapses \citep[for a review, see][]{Brette07_349}. This
update scheme restricts function calls of synapse code to time points
of spike events and thereby improves performance in simulations of
biologically plausible networks, where spike events at individual
synapses are rare and the total number of synapses is large compared
to the number of neurons. In complex voltage-based plasticity rules,
synapses, however, rely on continuous information of state variables
of postsynaptic cells to update their strength, which naturally suggests
their time-driven update. Instead, we here propose an efficient archiving
of voltage traces to enable event-based synapse updates and detail
two schemes for storage, read out and post-processing of time-continuous
or discontinuous information. We show their superior performance with
respect to time-driven update both theoretically and with a reference
implementation in the neural network simulation code NEST for the
rules proposed in \citet{Clopath10_344} and \citet{Urbanczik14}.

Event-driven update schemes for voltage-based plasticity come at the
expense of storing possibly long histories of a priori continuous
state variables. Such histories not only require space in memory but
they also affect the runtime of simulations, which we focus on here.
The time spent for searching and post-processing the history to calculate
weight updates increases with increasing length, and these operations
have to be done for each synapse. Therefore, in addition to an ordinary
event-driven scheme, we devised a compression scheme that becomes
superior for long histories as occurring in the Urbanczik-Senn rule.
In particular for networks with small in-degrees or synchronous spiking,
the compression scheme results in a shorter history. It further reduces
the total amount of computations for weight changes by partially re-using
results from other synapses thereby avoiding multiple processing of
the history. For short histories as occurring in the Clopath rule,
the compression results in unnecessary overhead and an increase in
history size as one entry per last presynaptic spike time needs to
be stored instead of a discontinuous membrane potential around sparse
postsynaptic spike events. We here, for simplicity, contrasted time-
and event-driven update schemes. However, further work could also
investigate hybrid schemes, where synapses are not only updated at
spike events, but also on a predefined and coarse time grid to avoid
long histories and corresponding extensive management. A similar mechanism
is used in \citet{Kunkel11_00160} to implement a normalization of
synaptic weights. The corresponding technical details can be found
in \citet[ch. 5.2]{Kunkel15}.

The storage and management of the history as well as complex weight
change computations naturally reduce the performance of simulations
with voltage-based plasticity in comparison to static or STDP synapses.
The latter only require information on spike times which is much less
data compared to continuous signals. Nevertheless, given that the
Clopath rule is based on thresholded membrane potentials and consequently
short, discontinuous histories, the performance and scaling of the
event-driven algorithms is only slightly worse than for ordinary STDP.
Time-driven implementations cannot employ this model feature and update
weights also in time steps where no adjustment would be required,
leading to significantly slower simulations. The performance gain
of using event-driven schemes is less pronounced for the Urbanczik-Senn
rule as, by design, histories are typically long. In this case, the
compression scheme naturally yields better results in terms of runtime.
Our own modification of the Urbanczik-Senn rule only requires storage
of sparsely sampled membrane potentials, giving rise to the same performance
as STDP. Generally, an algorithm is faster if it requires less computations.
However, opportunities for vectorization and cache efficient processing,
outside of the scope of the present manuscript, may change the picture.

We here chose the Clopath and the Urbanczik-Senn rule as two prototypical
models of voltage-based plasticity. While both rules describe a voltage
dependence of weight updates, their original motivation as well as
their specific form are different: The Clopath rule refines standard
STDP models to capture biologically observed phenomena such as frequency
dependence of weight changes \citep{Sjostrom01}. For this it is sufficient
to take into account membrane potential traces in the vicinity of
spike events, leading to storage of time-discontinuous histories in
our implementation. In contrast, the Urbanczik-Senn rule is functionally
inspired by segregating dendritic and somatic compartments of cells
and using the difference between somatic output and dendritic prediction
as a teacher signal for dendritic synapses. The teacher signal is
by construction never vanishing, imposing the need to store a time-continuous
history. The original publications of both rules had a great and long-lasting
impact on the field. The Clopath rule has been used in a variety of
studies \citep{Clopath10,Ko13_96,LitwinKumar14,Sadeh15_1,Bono17,Maes20_e1007606},
partly in modified versions which are, however, still compatible with
the here presented simulation algorithms. The same holds for the Urbanczik-Senn
rule \citep{Brea16_1,Sacramento18_8721}.

The current implementation, which is published and freely available
in NEST 2.20.1, supports an adaptive exponential integrate-and-fire
and a Hodgkin-Huxley neuron model for the Clopath rule. The former
is used in the original publication \citep{Clopath10_344} and the
latter appears on ModelDB \citep{Hines04_7} in code for the Clopath
rule for the NEURON simulator \citep{Hines01}. For the Urbanczik-Senn
rule, NEST currently supports the two-compartment Poisson model neuron
of the original publication \citep{Urbanczik14}. A three-compartment
version as used in \citet{Sacramento18_8721} or other models are
straight forward to integrate into the current simulation framework.
However, with voltage-based plasticity rules, borders between neurons
and synapses become blurred as these rules often depend on specifics
of the employed neuron models rather than only spike times as for
standard STDP. Consequently, archiving nodes might need to have specific
functionalities, which, in light of the zoo of existing neuron models,
could easily lead to a combinatorial explosion of code. These problems
can in future be overcome with automatic code generation using NESTML
that only creates and compiles code that is needed for the specified
model simulations \citep{Plotnikov_2016}.

While the here presented implementation refers to the neural network
simulator NEST \citep{Gewaltig_07_11204}, the proposed algorithms
and simulation infrastructure are compatible with any network simulator
with event-driven update of synapses, such as, for example, NEURON
\citep[cf. ch. 2.4]{Lytton2016_2063} and Brian2 \citep{Stimberg14}.
Furthermore, applicability is not restricted to the Clopath and Urbanczik-Senn
rule, but the framework can be adapted to any other learning rule
that relies on state variables of postsynaptic neurons. State variables
hereby not only encompass membrane potentials such as, for example,
in the LCP rule by \citet{Mayr10}, the Convallis rule by \citet{Yger13_1},
the voltage-triple rule by \citet{Brea13_9565}, the MPDP rule by
\citet{Albers16_1}, the voltage-gated learning rules by \citet{Brader07},
\citet{Sheik16_164}, \citet{Qiao15_141}, \citet{Diederich18_1}
and \citet{Cartiglia20_84}, or the branch-specific rule by \citet{Legenstein11_10787},
but also, for example, firing rates of stochastic neuron models or
rate models \citep{Brea16_1,Sacramento18_8721}, or other learning
signals \citep{Neftci17_324,Bellec19_738385}. The infrastructure
in NEST allows for the storage of time-continuous and discontinuous
histories and therefore poses no restrictions on the dependence of
the learning rule on the postsynaptic state variables. The here developed
machinery could be also used to store external teacher signals that
are provided to model neurons by stimulation devices mimicking brain
or environmental components not explicitly part of the model. Since
synapses are located at the compute process of the postsynaptic neuron,
readout of state variables from presynaptic neurons comes with large
costs in simulations on distributed computing architectures and is
therefore not considered here. Due to specifics of the present NEST
code in spike delivery, the event-driven compression proposed here
is only applicable in NEST for delays that equal the simulation time
step. Such a restriction can be readily overcome in a simulation algorithm
that performs a chronological update of synapses.

In general, one has to distinguish two types of efficiency in the
context of simulating plastic networks: Firstly, the biological time
it takes the network to learn a task by adapting the weights of connections.
Secondly, the wall-clock time it takes to simulate this learning process.
Both times crucially depend on the employed plasticity rule. In this
study, we focus on the wall-clock time and argue that this can be
optimized by designing learning rules that require storing only minimal
information on postsynaptic state variables. Ideally, the plasticity
rule contains unfiltered presynaptic or postsynaptic spike trains
to reach the same performance as in ordinary STDP simulations. If
rules, however, need to capture the pre- and post-spike dynamics of
postsynaptic neurons, it is beneficial to make use of thresholded
state variables as in the example of the Clopath rule as this yields
short, time-discontinuous histories. Reducing the amount of information
available for synapses to adjust their weights can in general slow
down the learning. We present a modification of the Urbanczik-Senn
rule where the dendritic prediction of the somatic firing contains
an additional sampling step with Poisson spike generation. This modification
significantly reduces the simulation time. For the here presented
simple task, learning speed is largely unaffected, but generally a
performance decrease is to be expected when error signals become more
noisy. Therefore, there is a trade-off between learning speed and
simulation speed, which should be considered in the design process
of new learning rules. \citet{Cartiglia20_84} propose another modification
of the Urbanczik-Senn rule underlying the model in \citet{Sacramento18_8721}:
this simplification only requires postsynaptic membrane potentials
at the time of spike events, which makes the rule much more efficient
to simulate and applicable to neuromorphic hardware. \citet{Bono17}
simplify the Clopath rule in an analogous fashion to allow for its
event-based simulation in the spiking network simulator Brian \citep{Stimberg14}.
Our general framework supports systematic testing of such simplifications
in terms of simulation performance and functionality.

For the plasticity rules by \citet{Clopath10_344} and \citet{Urbanczik14},
we present a highly scalable reference implementation that is published
and freely available in NEST 2.20.1. The parallelism of the NEST implementation
enables simulations of plastic networks of realistic size on biologically
plausible time scales for learning. The field of computational neuroscience
recently entered a new era with the development of large-scale network
models \citep{Markram2015_456,Schmidt18_e1006359,Billeh20}. Emulating
the dynamics of cortical networks, such models are so far restricted
to static connections. We here provide simulation algorithms for plasticity
mechanisms that are required for augmenting such complex models with
functionality. It is our hope that incorporating both biologically
and functionally inspired plasticity models in a single simulation
engine fosters the exchange of ideas between communities towards the
common goal of understanding system-level learning in the brain.

\section*{Code availability}

The reference implementation for the event-driven update scheme of
synapses with Clopath and Urbanczik-Senn plasticity was reviewed by
the NEST initiative and is publicly available in NEST 2.20.1. The
PyNEST code for model simulations and Python scripts for the analysis
and visualization of results are fully available at https://doi.org/10.5281/zenodo.4565188.

\section*{Conflict of Interest Statement\pdfbookmark[1]{Conflict of Interest Statement}{ConflictsPage}}

The authors declare that the research was conducted in the absence
of any commercial or financial relationships that could be construed
as a potential conflict of interest.

\section*{Acknowledgments\pdfbookmark[1]{Acknowledgments}{AcknowledgmentsPage}}

We thank Claudia Clopath and Wulfram Gerstner for explaining details
of their reference implementation and the underlying biological motivation.
Moreover, we thank Hedyeh Rezaei and Ad Aertsen for suggesting to
implement the Clopath rule in NEST and Charl Linssen, Alexander Seeholzer,
Renato Duarte for carefully reviewing our implementation. Finally
we thank Walter Senn, Mihai A. Petrovici, Laura Kriener and Jakob
Jordan for fruitful discussions on the Urbanczik-Senn rule and our
proposed spike based version. We further gratefully acknowledge the
computing time on the supercomputer JURECA \citep{stephan2015juqueen}
at Forschungszentrum J{\"u}lich granted by firstly the JARA-HPC Vergabegremium
(provided on the JARA-HPC partition, jinb33) and secondly by the Gauss
Centre for Supercomputing (GCS) (provided by the John von Neumann
Institute for Computing (NIC) on the GCS share, hwu12). This project
has received funding from the Helmholtz Association Initiative and
Networking Fund under project number SO-092 (Advanced Computing Architectures,
ACA) and the European Union's Horizon 2020 research and innovation
programme under grant agreements No 785907 (HBP SGA2) and No 945539
(HBP SGA3). Partly supported by the Helmholtz young investigator group
VH-NG-1028. All network simulations carried out with NEST (http://www.nest-simulator.org).
This work was supported by the J{\"u}lich-Aachen Research Alliance Center
for Simulation and Data Science (JARA-CSD) School for Simulation and
Data Science (SSD).

\section{Appendix}

\subsection{Analytical integration in Urbanczik-Senn rule}

\label{subsec:Analytical-integration-in}

To derive \prettyref{eq:us_weight_change} it is convenient to first
investigate $\Delta W_{ij}(0,t)$ and $\Delta W_{ij}(0,T)$ and then
compute $\Delta W_{ij}(t,T)=\Delta W_{ij}(0,T)-\Delta W_{ij}(0,t)$.
Assuming that the simulation starts at $t=0$, the weight change from
the start to time $t$ is given by
\begin{align*}
\Delta W_{ij}(0,t) & =\eta\int_{0}^{t}dt^{\prime}\,\int_{0}^{t^{\prime}}dt^{\prime\prime}\kappa\left(t^{\prime}-t^{\prime\prime}\right)V_{i}^{*}\left(t^{\prime\prime}\right)s_{j}^{*}\left(t^{\prime\prime}\right)\\
 & =\eta\int_{0}^{t}dt^{\prime\prime}\int_{t^{\prime\prime}}^{t}dt^{\prime}\,\kappa\left(t^{\prime}-t^{\prime\prime}\right)V_{i}^{*}\left(t^{\prime\prime}\right)s_{j}^{*}\left(t^{\prime\prime}\right)\\
 & =\eta\int_{0}^{t}dt^{\prime\prime}\,\left[\tilde{\kappa}\left(t-t^{\prime\prime}\right)-\tilde{\kappa}\left(0\right)\right]V_{i}^{*}\left(t^{\prime\prime}\right)s_{j}^{*}\left(t^{\prime\prime}\right)\\
 & =\eta\left[-I_{2}\left(0,t\right)+I_{1}\left(0,t\right)\right]
\end{align*}
where we exchanged the order of integration from the first to the
second line. In the third line we introduced $\tilde{\kappa}\left(t\right)$
defined by $\kappa\left(t\right)=\frac{\partial}{\partial t}\tilde{\kappa}\left(t\right)$
and in the fourth line we defined the two integrals
\begin{align*}
I_{1}\left(t_{1},t_{2}\right) & =-\int_{t_{1}}^{t_{2}}dt^{\prime}\,\tilde{\kappa}\left(0\right)V_{i}^{*}\left(t^{\prime}\right)s_{j}^{*}\left(t^{\prime}\right), & I_{2}\left(t_{1},t_{2}\right) & =-\int_{t_{1}}^{t_{2}}dt^{\prime}\,\tilde{\kappa}\left(t-t^{\prime}\right)V_{i}^{*}\left(t^{\prime}\right)s_{j}^{*}\left(t^{\prime}\right).
\end{align*}
In case of the Urbanczik-Senn rule
\[
\tilde{\kappa}\left(t\right)=-e^{-\frac{t}{\tau_{\kappa}}}
\]
which implies the identities
\begin{align*}
I_{1}\left(t_{1},t_{2}+\Delta t\right) & =I_{1}\left(t_{1},t_{2}\right)+I_{1}\left(t_{2},t_{2}+\Delta t\right), & I_{2}\left(t_{1},t_{2}+\Delta t\right) & =e^{-\frac{t_{2}-t_{1}}{\tau_{\kappa}}}I_{2}\left(t_{1},t_{2}\right)+I_{2}\left(t_{2},t_{2}+\Delta t\right),
\end{align*}
which we use to write the weight change from $t$ to $T$ as
\begin{align*}
\Delta W_{ij}(t,T) & =\Delta W_{ij}(0,T)-\Delta W_{ij}(0,t)\\
 & =\eta\left[-I_{2}\left(0,T\right)+I_{1}\left(0,T\right)+I_{2}\left(0,t\right)-I_{1}\left(0,t\right)\right]\\
 & =\eta\left[I_{1}\left(t,T\right)-I_{2}\left(t,T\right)+I_{2}\left(0,t\right)\left(1-e^{-\frac{T-t}{\tau_{\kappa}}}\right)\right].
\end{align*}
This is the the result \ref{eq:us_weight_change}.

\subsection{Voltage clamping of the adaptive exponential integrate-and-fire model}

\label{subsec:Voltage-clamping}

For the Clopath rule the change of the synaptic weight strongly depends
on the excursion of the membrane potential $V_{m}$ around a spike
of the postsynaptic neuron which causes $\bar{u}_{\pm}$ to cross
the respective thresholds $\theta_{\pm}$ so that \prettyref{eq:clopath_ltd}
and \prettyref{eq:clopath_ltp} yield nonvanishing results. Within
the original neuron model \citep{Brette-2005_3637} $u$ is reset
immediately after it reached the spiking threshold so that the shape
of the action potential is not described accurately. In our NEST implementation
of \texttt{aeif\_psc\_delta\_clopath} we adapted the approach of the
reference implementation on ModelDB \citep{Hines04_7} and introduced
a clamping of $u$ to a fixed value $V_{\mathrm{clamp}}$ for a period
of $t_{\mathrm{clamp}}$ before it is reset. The reference implementation
is restricted to a simulation resolution of exactly $1\,\mathrm{ms}$
and sets $u$ to two different values for the two subsequent simulation
steps after a spike. To be independent of the resolution of the simulation,
the implementation in NEST uses a constant $V_{\mathrm{clamp}}$.
In the simulations we set $t_{\mathrm{clamp}}$ to $2\,\mathrm{ms}$
and $V_{\mathrm{clamp}}$ to $33\,\mathrm{mV}$. These values are
chosen to match the behavior of the reference implementation.

\subsection{History management \label{subsec:History-management}}

There are three points that need to be considered in the context of
history management: First, which information needs to be stored.
Second, how to search and read out the history. Third, how to identify
and remove information that is no longer needed. The first and third
point mainly affect memory usage, while the second point impacts the
simulation time as searching within shorter histories is faster.

There are four different histories to which our considerations apply.
The one to store the membrane potential $V_{i}^{*}$, the compressed
history $\Delta W_{i}(t_{LS},T)$ used only for the compressed event-driven
scheme, the history to store the spike times $s_{i}$ of the postsynaptic
neuron (also used for ordinary STDP), and finally one might need a
history that stores the last spike time for every incoming synapse
(see below for details).

\paragraph{Adding information to the history:}

This paragraph concerns only the history that stores the time trace
of $V_{i}^{*}$. In every time step of the simulation, neurons call
the protected function \texttt{write\_history()} of the archiving
node and pass the current value of the (low-pass filtered) membrane
potential. The archiving node then computes the derived quantities
$V_{i}^{*}$ or  combinations of $V_{i}^{*}$ and $s_{i}^{*}$, and
saves them in the history, which is of type \texttt{vector}. It is
more efficient to do the computations inside the archiving node and
not in the synapse for two reasons: Firstly, the computation is done
only once and then used for all incoming synapses. This way no direct
exchange of information between different synapses is required. Secondly,
the archiving node does not need to store the raw membrane potentials
before readout, but can directly store the derived quantities $V_{i}^{*}$,
which reduces the memory consumption, especially in cases where only
a non-continuous history is needed.

\paragraph{Readout of information from the history:}

Let's assume $t_{\mathrm{LS}}$ and $t_{\mathrm{S}}$ be the times
of the last and the current spike of a synapse. At time $t_{\mathrm{S}}$
that synapse then needs to request a part from $t_{1}=t_{\mathrm{LS}}-d$
to $t_{2}=t_{\mathrm{S}}-d>t_{1}$ of the history that ranges from
$t_{\mathrm{start}}<t_{1}$ to $t_{\mathrm{end}}>t_{2}$. This part
is shifted with respect to the spike times by a delay $d$ which models
the time of signal propagation from the postsynaptic soma back to
the synapse. The software framework NEST of our reference implementation
uses only one variable to represent the delay from synapse to soma
and the delay in the opposite direction. Consequently, when a spike
arrives at the synapse of the postsynaptic neuron, the synapse sees
a membrane potential from the past. In case every time step of the
simulation adds a new entry to the history, one can easily compute
the positions of the entries corresponding to $t_{1/2}$ by just knowing
$t_{\mathrm{start}}$ and $t_{\mathrm{end}}$. As pointed out in \ref{subsec:Reference-implementation}
this is the case for the Urbanczik-Senn plasticity rule. If the history
is not continuous in time, like in case of the Clopath rule, this
scheme is not applicable. Instead, we add a time stamp $s$ as an
additional variable to each entry and search for those with the smallest/greatest
$s$ within the interval $(t_{1},t_{2})$ using e.g. a linear or a
binary search. Searching for the positions that define the start and
the end of the requested interval is slower than computing them directly.
Nevertheless, a non-continuous history can lead to a large acceleration
of simulations  as we discussed in case of the Clopath rule (\prettyref{subsec:Clopath-performance}).
Here, only values of the membrane potential in the vicinity of a spike
of the postsynaptic neuron are needed so that neglecting the majority
of values in between leads to a non-continuous history but saves memory.

Technically, the archiving node contains a function called \texttt{get\_history()}
which expects two iterators \texttt{start} and \texttt{finish} and
two times $t_{1}$ and $t_{2}$. When executed, the function sets
the iterators to point to the correct entries of the history of the
postsynaptic neuron corresponding to $t_{1}$ and $t_{2}$, respectively.
Having received the correct position of the pointers, the synapse
evaluates the integral \prettyref{eq:int_plasticity-1}. In the event-driven
compression scheme, the integration \prettyref{eq:Delta W_i} is not
done inside the synapse but inside the \texttt{archiving\_node}. The
reason for this is that the compressed history $\Delta W_{i}(t_{LS},t_{\mathrm{S}})$,
which is updated in case of an incoming spike, is stored inside the
\texttt{archiving\_node.} This way no exchange of information is needed.
The synapse only triggers the updating process by calling the function
\texttt{compress\_history()} of the \texttt{archiving\_node}. Internally,
the \texttt{archiving\_node} can use \texttt{get\_history()} to obtain
the part of the history that has to be integrated. Even though the
linear search a priori might seem less efficient than a binary search
or direct computation of the position, it turns out that it has an
advantage in that it iterates consecutively over the history entries
which can be employed to identify data no longer needed. Therefore,
especially for short histories a simple iteration that comes without
any overhead is fastest (see \prettyref{subsec:Clopath-performance}).

\paragraph{Removing information from the history:}

To prevent the history from occupying an unnecessary amount of memory,
it is crucial to have a mechanism to delete those entries that have
been used by all incoming synapses. The simplest implementation to
identify these entries is to add one additional variable to each entry
called \textit{access counter} initialized to zero when the entry
is created. When a synapse requests a part from $t_{1}$ to $t_{2}$
of the history, the algorithm iterates over all entries $t_{1}<t<t_{2}$
and increases the access counters by one. After the update of the
synaptic weight all entries whose access counters are equal to the
number of incoming synapses are deleted. This scheme can be combined
easily with a linear search starting the iteration from the oldest
entry of the history.

For long histories a linear search is inefficient and should be replaced
by a binary search or direct computation of positions if applicable.
To avoid iteration within long histories, we replace access counters
by a vector that stores the last spike time $t_{\mathrm{LS}}$ for
every incoming synapse. If a synapse delivers a spike, it updates
its entry in that vector by replacing $t_{\mathrm{LS}}$ by the time
stamp of the current spike. After each weight update, searching the
vector for the smallest $t_{\mathrm{LS}}$ allows us to safely remove
all membrane potentials with time stamps $t<\min(\{t_{\mathrm{LS},i}\})$.
In practice, we can further improve this mechanism with two technical
details. Firstly, $n$ incoming spikes with the same time stamp can
share the same entry $t_{\mathrm{LS}}$ which we then have to provide
with a counter that goes down from $n$ to zero in steps of one whenever
one of the $n$ synapses sends a new spike for a time $t>t_{LS}$.
Secondly, we can avoid the search for the smallest $t_{LS}$ by making
sure that the entries $t_{\mathrm{LS}}$ are in chronological order.
This can be easily achieved if a synapse does not update its entry
in the vector but removes it and appends a new one at the end of the
vector.

\subsection{Implementation of experiments using Clopath rule}

\label{subsec:Implementation-of-experiments}

\subsubsection{Spike pairing experiment}

\label{subsec:Spike-pairing-experiment}

The setup of the spike pairing experiment from \citet{Clopath10_344}
presented in \prettyref{fig:clopath_results}A,B consists of two neurons
connected via a plastic synapse. The pre- and postsynaptic neuron
are forced to spike with a time delay of $\Delta t$ by stimulation
with \texttt{spike\_generators} at times $t_{\mathrm{pre}}^{\left(i\right)}=t^{\left(i\right)}$
and $t_{\mathrm{post}}^{\left(i\right)}=t^{\left(i\right)}+\Delta t$,
respectively. A positive time shift $\Delta t>0$ refers to the presynaptic
neuron spike before the postsynaptic one (pre-post pairing, solid
lines in \prettyref{fig:clopath_results}) and vice versa. Spike pairs
$\left(t_{\mathrm{pre}}^{\left(i\right)},t_{\mathrm{post}}^{\left(i\right)}\right)$
are induced with frequency $f_{\mathrm{pair}}=\frac{1}{t^{(i+1)}-t^{(i)}}$
and the weight change of the synapse is measured after a set of five
pairs. In our simulation using NEST the presynaptic neuron is modeled
as a \texttt{parrot\_neuron} and the postsynaptic neuron is either
of type \texttt{aeif\_psc\_delta\_clopath} or \texttt{hh\_psc\_alpha\_clopath}.
In NEST \texttt{parrot\_neurons} are model neurons that emit a spike
whenever they receive one. In this setup they are required because
devices like \texttt{spike\_generators} support only static synapses
in NEST so that we cannot connect the postsynaptic neuron directly
to the \texttt{spike\_generator} via a plastic synapse. The initial
weight of the \texttt{clopath\_synapse} connecting the two neurons
is given by $w_{\mathrm{init}}$. In this experiment we use the Clopath
rule with fixed amplitude $A_{\mathrm{LTD}}$. A list with all the
parameters can be found in Table S1 in supplement.

\subsubsection{Emergence of strong bidirectional couplings}

In this experiment after \citet{Clopath10_344}, a small network of
$N_{I}$ inhibitory and $N_{E}$ excitatory neurons subject to an
external input develops strong bidirectional couplings between neurons
of the excitatory population. The input is given by $N_{\mathrm{p}}$
Poisson spike trains with rates 
\begin{equation}
f_{\mathrm{p}}^{\left(j\right)}=A_{\mathrm{p}}e^{-\frac{\left(j-\mu_{\mathrm{p}}\right)^{2}}{2\sigma_{\mathrm{p}}^{2}}}+c_{\mathrm{p}},\label{eq:clopath_network_rate}
\end{equation}
where $j=1,\ldots,N_{\mathrm{p}}$. The center $\mu_{\mathrm{p}}$
of the Gaussian is drawn randomly from a set $s_{\mathrm{p}}$ of
possible values and a new value is drawn after each time interval
$t_{\mu}$. The total number of intervals is $N_{\mu}$. In our simulation
with NEST we used \texttt{aeif\_psc\_delta} model neurons with the
same parameters (cf. Table S3 in supplement) for both the inhibitory
and the excitatory population. The simulation is divided into $N_{\mu}$
intervals between which the rates of the $N_{\mathrm{p}}$ \texttt{poisson\_generators}
are set according to \prettyref{eq:clopath_network_rate}. The \texttt{poisson\_generators}
are connected in a one-to-one manner to $N_{\mathrm{p}}$ \texttt{parrot\_neurons}
which in turn are connected to the network. The details of the latter
connectivity can be found in Table S2 in supplement. In NEST a \texttt{poisson\_generator}
that is connected to several target model neurons generates an independent
Poisson spike train for each of these neurons. Thus, the intermediate
step via \texttt{parrot\_neurons} is required to provide neurons in
the network with common Poisson inputs. Moreover, a direct connection
from a device like a \texttt{poisson\_generator} to a model neuron
via a plastic synapse is not possible in NEST. In this experiment,
the Clopath rule with homeostasis (time dependent prefactor for LTD,
cf. \prettyref{subsec:Implementation-ubarbar}) is used. \prettyref{fig:clopath_results}
C shows the weights of the synapses among the excitatory population
after the simulation.

\subsection{Implementation of homeostasis $A_{\mathrm{LTD}}\left(\bar{\bar{u}}\right)$}

\label{subsec:Implementation-ubarbar}

For the network simulations presented in \citet{Clopath10_344}, the
authors use a sightly modified version of the Clopath rule defined
in \prettyref{eq:clopath_ltd}: The constant factor $A_{\mathrm{LTD}}$
is replaced by a voltage dependent term 
\[
A_{\mathrm{LTD}}\left(\bar{\bar{u}}\right)=A_{\mathrm{LTD}}\left(\frac{\bar{\bar{u}}}{u_{\mathrm{ref}}}\right)^{2}
\]
to take into account homeostatic processes. The quantity $\bar{\bar{u}}$
is a temporal average of the quantity $\bar{u}_{-}\left(t\right)$
over a time window of $T=1\,\,s$ and $u_{\mathrm{ref}}$ is a reference
value. An exact temporal average requires storing the time trace of
$\bar{u}_{-}\left(t\right)$ for the entire interval $T$. This would
cancel the advantage of keeping only a sparse history as discussed
in \prettyref{par:clopath_specifics_ed} where storage of time traces
is needed only in the vicinity of spikes. Therefore, deviating from
the original work by \citet{Clopath10_344}, we implement an additional
low-pass filtering $\bar{\bar{u}}\left(t\right)=\left(\kappa_{\mathrm{low}}\ast\bar{u}_{-}\right)\left(t\right)$
with an exponential kernel $\kappa_{\mathrm{low}}\left(t\right)=H\left(t\right)\exp\left(-t/\bar{\bar{\tau}}\right)$
instead. Like $\bar{u}_{\pm}$, $\bar{\bar{u}}$ is passed as an additional
state variable to the solver.

\subsection{Implementation of experiment using Urbanczik-Senn rule}

\label{subsec:Implementation--US}

In the simulation experiment shown in \prettyref{fig:US-reproduction}
the dendrite of a conductance-based two-compartment model neuron receives
a spike pattern of duration $T$ as an input via plastic synapses.
The pattern consists of $N_{\mathrm{p}}$ independent Poisson spike
trains with a firing rate $f_{\mathrm{p}}$. For learning, the pattern
is repeated $N_{\mathrm{rep}}$ times. Dendritic synapses adapt their
weights so that after learning the somatic membrane potential $U$
and the dendritic prediction $V_{w}^{\ast}$ follow a matching potential
$U_{\mathrm{M}}$. The latter is created by somatic input via two
\texttt{spike\_generators} that are connected via a static excitatory
or inhibitory connection, respectively. Both spike generators send
spikes in every simulation step. Inhibitory input spikes have a constant
weight to generate a constant somatic inhibitory conductance $g_{I}$.
Excitatory spikes have a modulated weight to generate a periodic excitatory
conductance $g_{E}$. The input to the dendritic compartment is provided
by $N_{\mathrm{p}}$ \texttt{spike\_generators} each of which is connected
to one \texttt{parrot\_neuron} which in turn is connected to the dendrite
via a plastic \texttt{urbanczik\_synapse}. The intermediate parrot\_neurons
are required since in NEST the spike\_generators can have only static
synapses as outgoing connections. The spike times of the \texttt{spike\_generators}
are set to repeatedly generate the spike pattern created before the
start of the actual simulation. The neuron's state variables are read
out by a \texttt{multimeter} and the synaptic weights by a \texttt{weight\_recorder}.

\subsection{Experiment with modified version of the Urbanczik-Senn rule}

\label{subsec:US_experiment_modified}

\begin{figure}[H]
  \centering{}\includegraphics[width=0.75\paperwidth]{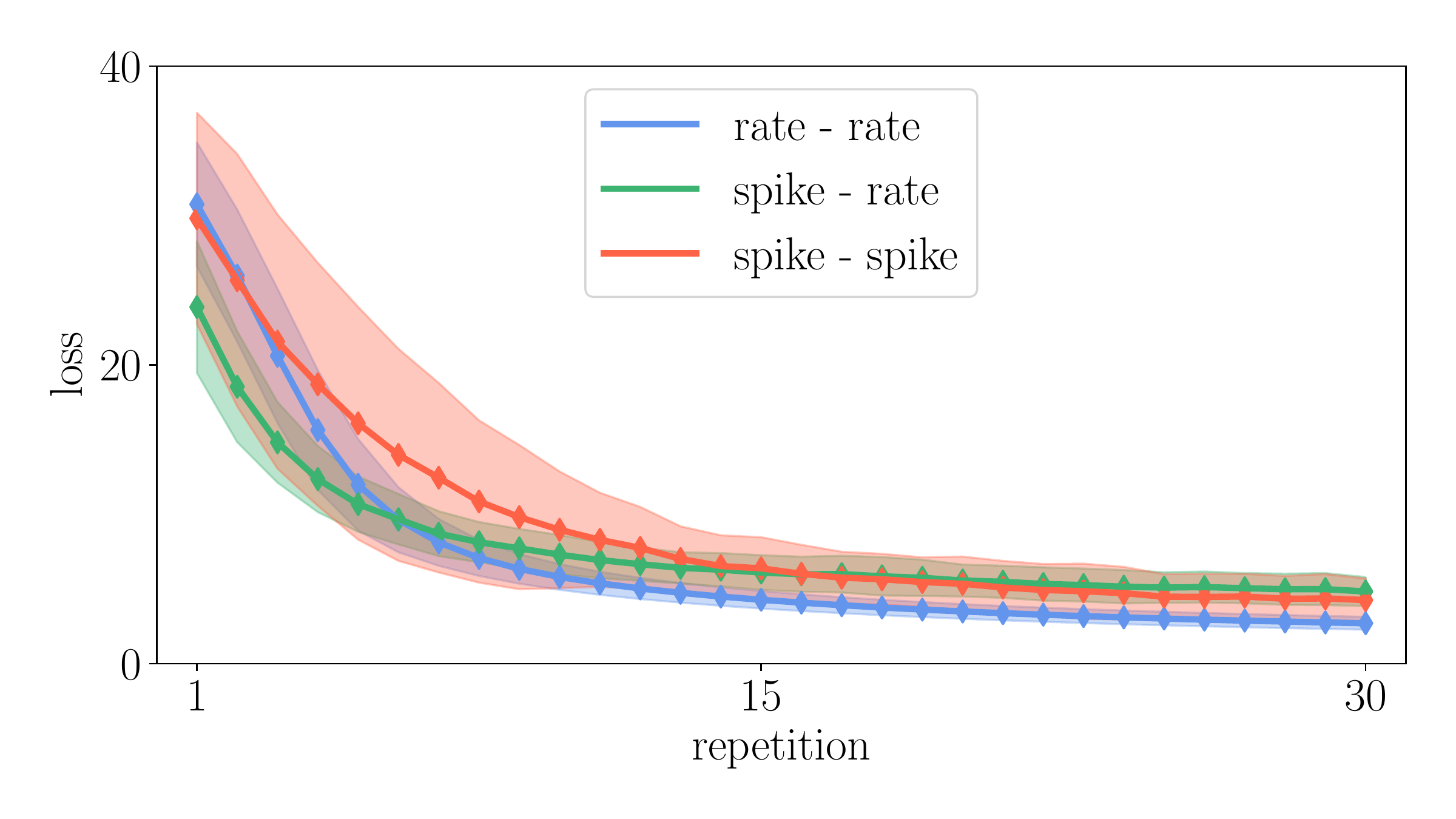}\caption{\textbf{Comparison of learning curves in the experiment described
in \prettyref{subsec:Implementation--US} for different variants of
the Urbanczik-Senn plasticity rule}. The loss is averaged over $128$
trials of different input patterns. Solid curves denote the mean value
and the shaded area the corresponding standard deviation of the loss.\label{fig:US_modified}}
\end{figure}

The weight change of the Urbanczik-Senn rule as presented in \prettyref{subsec:Example_US}
in line with the original publication is driven by the prediction
error
\[
V_{i}^{*}=\left(s_{i}-\phi(V_{i})\right)\,h\left(V_{i}\right),
\]
where $s_{i}$ is the somatic spike train and $V_{i}$ the dendritic
prediction of the somatic membrane potential $U_{i}$. Instead of
integrating over the difference between the spike train and the rate
$\phi\left(V_{i}\right)$ (spike-rate), one can derive two variants
\begin{align*}
V_{i}^{*}= & \left(s_{i}-s_{i}^{\mathrm{dend}}\right)\,h\left(V_{i}\right)\quad\mathrm{(spike-spike)}\quad\mathrm{and}\\
V_{i}^{*}= & \left(\phi\left(U_{i}\right)-\phi\left(V_{i}\right)\right)\,h\left(V_{i}\right)\quad\mathrm{(rate-rate)}.
\end{align*}
In the first one (spike-spike) we replaced the dendritic rate prediction
by a noisy realization $s_{i}^{\mathrm{dend}}$ using an inhomogeneous
Poisson process with rate $\phi\left(V_{i}\right)$. In the second
one (rate-rate) the somatic spike train is replaced by the rate of
the underlying Poisson process which is computed by applying the rate
function $\phi$ to the somatic potential $U_{i}$. The learning of
a matching potential $U_{M}$ as described in \prettyref{subsec:Reproduction-of-results}
also works in these two cases. \prettyref{fig:US_modified} shows
the learning curve for all three variants of the Urbanczik-Senn rule.
The loss is defined as the average mismatch between $U_{i}$ and $U_{M}$
averaged over one period $T_{p}$ of the input pattern
\[
\frac{1}{T_{p}}\int dt\left(U\left(t\right)-U_{M}\left(t\right)\right)^{2}.
\]
The decrease of the loss as a function of the pattern repetitions
has a similar shape for all three variants with a significantly higher
variance in case of the spike-spike version.

\newpage{}

\section{Supplemental Material}

\renewcommand{\thetable}{S\arabic{table}}
\setcounter{table}{0}

\begin{table}[H]
\begin{raggedright}
\textcolor{black}{}%
\begin{tabular}{@{\hspace*{1mm}}p{3cm}@{}|@{\hspace*{1mm}}p{3.8cm}|@{\hspace*{1mm}}p{8.3cm}}
\hline 
\multicolumn{3}{c}{\textbf{A: Simulation parameters}}\tabularnewline
\hline 
\textbf{\textcolor{black}{Symbol}} & \textbf{\textcolor{black}{Value}} & \textbf{Description}\tabularnewline
$f_{\mathrm{pair}}$ & $\left[10,\,11,\ldots,50\right]\,\mathrm{Hz}$ & frequency of occurrence of spike pairs\tabularnewline
$\Delta t$ & $\pm10\,\mathrm{ms}$ & time shift of spike pair\tabularnewline
$w_{\mathrm{init}}\,\mathrm{[mV]}$ & $0.5$ $\mathrm{mV}$ or $\mathrm{pA/ms}$ & initial weight (unit is $\mathrm{mV}$ for aeif and $\mathrm{pA/ms}$
for hh neuron)\tabularnewline
\hline 
\end{tabular}
\par\end{raggedright}
\begin{raggedright}
\textcolor{black}{}%
\begin{tabular}{@{\hspace*{1mm}}p{3cm}@{}|@{\hspace*{1mm}}p{3.8cm}|@{\hspace*{1mm}}p{8.3cm}}
\multicolumn{3}{c}{\textbf{B: Parameters of }\texttt{\textbf{aeif\_psc\_delta\_clopath}}}\tabularnewline
\hline 
\textbf{\textcolor{black}{Symbol}} & \textbf{\textcolor{black}{Value}} & \textbf{Description}\tabularnewline
$E_{L}$ & $-54.402\,\mathrm{mV}$ & leak reversal potential\tabularnewline
$E_{\mathrm{Na}}$ & $50.0\,\mathrm{mV}$ & sodium reversal potential\tabularnewline
$E_{\mathrm{K}}$ & $-77.0\,\mathrm{mV}$ & potassium reversal potential\tabularnewline
$g_{L}$ & $30.0\,\mathrm{nS}$ & leak conductance\tabularnewline
$g_{\mathrm{Na}}$ & $12\cdot10^{3}\,\mathrm{nS}$ & sodium peak conductance\tabularnewline
$g_{\mathrm{K}}$ & $3.6\cdot10^{3}\,\mathrm{nS}$ & potassium peak conductance\tabularnewline
$C_{m}$ & $100\,\mathrm{pF}$ & membrane capacitance\tabularnewline
$\tau_{\mathrm{ex}}$ & $0.2\,\mathrm{ms}$ & rise time of the exc. synaptic alpha funct.\tabularnewline
$\tau_{\mathrm{in}}$ & $2.0\,\mathrm{ms}$ & rise time of the inh. synaptic alpha funct.\tabularnewline
$\theta_{-}$ & $-64.9\,\mathrm{mV}$ & threshold\tabularnewline
$\theta_{+}$ & $-35\,\mathrm{mV}$ & threshold\tabularnewline
$A_{\mathrm{LTD}}$ & $1\cdot10^{-4}\,\mathrm{1/mV}$ & amplitude of LTD\tabularnewline
$A_{\mathrm{LTP}}$ & $12\cdot10^{-4}\,1/\mathrm{mV}^{2}$ & amplitude of LTP\tabularnewline
$\tau_{-}$ & $10\,\mathrm{ms}$ & time constant of $\bar{u}_{-}$\tabularnewline
$\tau_{+}$ & $114\,\mathrm{ms}$ & time constant of $\bar{u}_{+}$\tabularnewline
$\tau_{s}$ & $15\,\mathrm{ms}$ & time constant of $s_{j}^{*}$\tabularnewline
$d_{s}$ & $5\,\mathrm{ms}$ & delay of $\bar{u}_{\pm}$\tabularnewline
\hline 
\end{tabular}
\par\end{raggedright}
\begin{raggedright}
\textcolor{black}{}%
\begin{tabular}{@{\hspace*{1mm}}p{3cm}@{}|@{\hspace*{1mm}}p{3.8cm}|@{\hspace*{1mm}}p{8.3cm}}
\multicolumn{3}{c}{\textbf{C: Parameters of }\texttt{\textbf{hh\_psc\_alpha\_clopath}}}\tabularnewline
\hline 
\textbf{\textcolor{black}{Symbol}} & \textbf{\textcolor{black}{Value}} & \textbf{Description}\tabularnewline
$E_{L}$ & $-70.6\,\mathrm{mV}$ & leak reversal potential\tabularnewline
$g_{L}$ & $30\,\mathrm{nS}$ & leak conductance\tabularnewline
$C_{m}$ & $281\,\mathrm{pF}$ & membrane capacitance\tabularnewline
$V_{\mathrm{reset}}$ & $-60\,\mathrm{mV}$ & reset value of membr. pot. after spike\tabularnewline
$V_{\mathrm{peak}}$ & $33\,\mathrm{mV}$ & spike detection threshold\tabularnewline
$\Delta_{T}$ & $2\,\mathrm{mV}$ & slope factor\tabularnewline
$\tau_{w}$ & $144\,\mathrm{ms}$ & spike adaptation time constant\tabularnewline
$\tau_{z}$ & $40\,\mathrm{ms}$ & spike adaptation time constant\tabularnewline
$V_{\mathrm{th,max}}$ & $30.4\,\mathrm{mV}$ & threshold potential after spike\tabularnewline
$\tau_{V,\mathrm{th}}$ & $50\,\mathrm{ms}$ & threshold potential time constant\tabularnewline
$a$ & $4\,\mathrm{nS}$ & subthreshold adaptation\tabularnewline
$b$ & $0.0805\,\mathrm{pA}$ & spike triggered adaptation\tabularnewline
$\theta_{-}$ & $-70.6\,\mathrm{mV}$ & threshold of $\bar{u}_{-}$\tabularnewline
$\theta_{+}$ & $-45.3\,\mathrm{mV}$ & threshold of $\bar{u}_{+}$\tabularnewline
$A_{\mathrm{LTD}}$ & $14\cdot10^{-5}\,\mathrm{1/mV}$ & amplitude of LTD\tabularnewline
$A_{\mathrm{LTP}}$ & $8\cdot10^{-5}\,1/\mathrm{mV}^{2}$ & amplitude of LTP\tabularnewline
$\tau_{-}$ & $10\,\mathrm{ms}$ & time constant of $\bar{u}_{-}$\tabularnewline
$\tau_{+}$ & $7\,\mathrm{ms}$ & time constant of $\bar{u}_{+}$\tabularnewline
$\tau_{s}$ & $15\,\mathrm{ms}$ & time constant of $s_{j}^{*}$\tabularnewline
$d_{s}$ & $4\,\mathrm{ms}$ & delay of $\bar{u}_{\pm}$\tabularnewline
\hline 
\end{tabular}
\par\end{raggedright}
\caption{\textbf{Parameters of the spike pairing experiment using the Clopath
rule.} The values for the aeif model are taken from \citep[Tab. 1 and appendix]{Clopath10_344}
and those for the hh model are extracted from the reference implementation
by B. Torben-Nielson on ModelDB \citep{Hines04_7}. \label{tab:spike_pairing_params-1}}
\end{table}

\newpage{}

\textcolor{black}{}
\begin{table}[H]
\begin{raggedright}
\textcolor{black}{}%
\begin{tabular}{@{\hspace*{1mm}}p{3cm}@{}|@{\hspace*{1mm}}p{12.2cm}}
\hline 
\multicolumn{2}{>{\centering}m{15cm}}{\textbf{\textcolor{black}{A: Model summary}}}\tabularnewline
\hline 
\textbf{\textcolor{black}{Populations}} & \textcolor{black}{Three: excitatory, inhibitory, external input}\tabularnewline
\textbf{\textcolor{black}{Connectivity}} & \textcolor{black}{all-to-all, fixed out-degree, fixed in-degree}\tabularnewline
\textbf{\textcolor{black}{Neuron model}} & adaptive exponential integrate-and-fire (aeif, Clopath)\tabularnewline
\textbf{\textcolor{black}{Plasticity}} & Clopath synapse\tabularnewline
\textbf{\textcolor{black}{Input}} & \textcolor{black}{independent homogeneous Poisson spike trains}\tabularnewline
\textbf{\textcolor{black}{Measurements}} & synapse weight\tabularnewline
\hline 
\end{tabular}
\par\end{raggedright}
\begin{raggedright}
\textcolor{black}{}%
\begin{tabular}{@{\hspace*{1mm}}p{3cm}@{}|@{\hspace*{1mm}}p{5.6cm}@{}|@{\hspace*{1mm}}p{6.6cm}}
\multicolumn{3}{>{\centering}m{15cm}}{\textbf{\textcolor{black}{B: Populations}}}\tabularnewline
\hline 
\textbf{\textcolor{black}{Name}} & \textbf{\textcolor{black}{Elements}} & \textbf{\textcolor{black}{Population size}}\tabularnewline
\textcolor{black}{E} & aeif/two-comp. & $N_{E}=10$\tabularnewline
I & aeif/two-comp. & $N_{I}=3$\tabularnewline
E\textsubscript{ext} & Poisson generator & $N_{p}=500$\tabularnewline
\hline 
\end{tabular}
\par\end{raggedright}
\begin{raggedright}
\textcolor{black}{}%
\begin{tabular}{@{\hspace*{1mm}}p{3cm}@{}|@{\hspace*{1mm}}p{2.8cm}@{}|@{\hspace*{1mm}}p{2.8cm}@{}|@{\hspace*{1mm}}p{6.6cm}@{}}
\multicolumn{4}{>{\centering}p{15cm}}{\textbf{\textcolor{black}{C: Connectivity}}}\tabularnewline
\hline 
\textbf{Name} & \textbf{\textcolor{black}{Source}} & \textbf{Target} & \textbf{Pattern}\tabularnewline
ExcExc & E & E & all-to-all (no autapses)\tabularnewline
ExcInh & E & I & fixed in-degree $C_{E}=8$\tabularnewline
InhExc & I & E & fixed out-degree $C_{I}=6$\tabularnewline
ExtExc & E\textsubscript{Ext} & E & all-to-all\tabularnewline
ExtInh & E\textsubscript{Ext} & I & all-to-all\tabularnewline
\hline 
\end{tabular}
\par\end{raggedright}
\begin{raggedright}
\textcolor{black}{}%
\begin{tabular}{@{\hspace*{1mm}}p{3cm}@{}|@{\hspace*{1mm}}p{12.4cm}@{}}
\multicolumn{2}{>{\centering}m{15cm}}{\textbf{\textcolor{black}{D: Neurons}}}\tabularnewline
\hline 
\textbf{\textcolor{black}{Name}} & \texttt{aeif\_psc\_delta\_clopath}\tabularnewline
\textbf{\textcolor{black}{Type}} & adaptive exponential integrate-and-fire\tabularnewline
\textbf{Details} & see \citet{Clopath10_344}\tabularnewline
\textbf{Parameters} & see \prettyref{tab:neuron_params_cl_network-1}\tabularnewline
\hline 
\end{tabular}
\par\end{raggedright}
\begin{raggedright}
\textcolor{black}{}%
\begin{tabular}{@{\hspace*{1mm}}p{3cm}@{}|@{\hspace*{1mm}}p{3.9cm}@{}|@{\hspace*{1mm}}p{5.5cm}@{}|@{\hspace*{1mm}}p{2.8cm}@{}}
\multicolumn{4}{>{\centering}p{15cm}}{\textbf{\textcolor{black}{E: Synapses}}}\tabularnewline
\hline 
\textbf{Name} & \textbf{Model} & \textbf{Initial weight {[}mV{]}} & \textbf{Max. weight {[}mV{]}}\tabularnewline
ExcExc & \texttt{clopath\_synapse} & $0.25$ & $0.75$\tabularnewline
ExcInh & \texttt{static\_synapse} & $1.0$ & $-$\tabularnewline
InhExc & \texttt{static\_synapse} & $1.0$ & $-$\tabularnewline
ExtExc & \texttt{clopath\_synapse} & random uniform from $\left[0.5,1.5\right]$ & $3.0$\tabularnewline
ExtInh & \texttt{static\_synapse} & random uniform from $\left[0.0,0.5\right]$ & $-$\tabularnewline
\hline 
\end{tabular}
\par\end{raggedright}
\begin{raggedright}
\textcolor{black}{}%
\begin{tabular}{@{\hspace*{1mm}}p{3cm}@{}|@{\hspace*{1mm}}p{12.2cm}}
\multicolumn{2}{>{\centering}m{15cm}}{\textbf{\textcolor{black}{F: Input}}}\tabularnewline
\hline 
\textbf{\textcolor{black}{Type}} & \textcolor{black}{Poisson generator}\tabularnewline
\textbf{\textcolor{black}{Description}} & \textcolor{black}{homogeneous Poisson spike trains, independent for
each neuron, modulated rate}\tabularnewline
\textbf{Parameters} & see \prettyref{tab:neuron_params_cl_network-1}\tabularnewline
\hline 
\end{tabular}
\par\end{raggedright}
\raggedright{}\textcolor{black}{\caption{\textbf{Model description of a small excitatory-inhibitory network
after }\citet{Nordlie-2009_e1000456}. This network reproduces the
emergence of strong bidirectional couplings using the Clopath rule
shown in Figure 7 . The values of the parameters are shown in \prettyref{tab:neuron_params_cl_network-1}.
\label{tab:Network-and-simulation-1-1}}
}
\end{table}

\newpage{}

\begin{table}[H]
\begin{raggedright}
\textcolor{black}{}%
\begin{tabular}{@{\hspace*{1mm}}p{3cm}@{}|@{\hspace*{1mm}}p{2.8cm}|@{\hspace*{1mm}}p{9.3cm}}
\hline 
\multicolumn{3}{c}{\textbf{A: Parameters of }\texttt{\textbf{aeif\_psc\_delta\_clopath}}}\tabularnewline
\hline 
\textbf{\textcolor{black}{Symbol}} & \textbf{\textcolor{black}{Value}} & \textbf{Description}\tabularnewline
$E_{L}$ & $-70.6\,\mathrm{mV}$ & leak reversal potential\tabularnewline
$g_{L}$ & $30\,\mathrm{nS}$ & leak conductance\tabularnewline
$C_{m}$ & $281\,\mathrm{pF}$ & membrane capacitance\tabularnewline
$V_{\mathrm{reset}}$ & $-60\,\mathrm{mV}$ & reset value of membr. pot. after spike\tabularnewline
$V_{\mathrm{peak}}$ & $20\,\mathrm{mV}$ & spike detection threshold\tabularnewline
$\Delta_{T}$ & $2\,\mathrm{mV}$ & slope factor\tabularnewline
$\tau_{w}$ & $144\,\mathrm{ms}$ & spike adaptation time constant\tabularnewline
$\tau_{z}$ & $40\,\mathrm{ms}$ & spike adaptation time constant\tabularnewline
$V_{\mathrm{th,max}}$ & $30.4\,\mathrm{mV}$ & threshold potential after spike\tabularnewline
$\tau_{V,\mathrm{th}}$ & $50\,\mathrm{ms}$ & threshold potential time constant\tabularnewline
$a$ & $4\,\mathrm{nS}$ & subthreshold adaptation\tabularnewline
$b$ & $0.0805\,\mathrm{pA}$ & spike triggered adaptation\tabularnewline
$\theta_{-}$ & $-70.6\,\mathrm{mV}$ & threshold of $\bar{u}_{-}$\tabularnewline
$\theta_{+}$ & $-45.3\,\mathrm{mV}$ & threshold of $\bar{u}_{+}$\tabularnewline
$A_{\mathrm{LTD}}$ & $14\cdot10^{-5}\,\mathrm{1/mV}$ & amplitude of LTD\tabularnewline
$u_{\mathrm{ref}}$ & $60$ & reference value for $\bar{\bar{u}}$\tabularnewline
$\bar{\bar{\tau}}$ & $1.5\cdot10^{3}\,\mathrm{ms}$ & time constant of $\bar{\bar{u}}$\tabularnewline
$A_{\mathrm{LTP}}$ & $8\cdot10^{-5}\,1/\mathrm{mV}^{2}$ & amplitude of LTP\tabularnewline
$\tau_{-}$ & $10\,\mathrm{ms}$ & time constant of $\bar{u}_{-}$\tabularnewline
$\tau_{+}$ & $7\,\mathrm{ms}$ & time constant of $\bar{u}_{+}$\tabularnewline
$\tau_{s}$ & $15\,\mathrm{ms}$ & time constant of $s_{j}^{*}$\tabularnewline
$d_{s}$ & $4\,\mathrm{ms}$ & delay of $\bar{u}_{\pm}$\tabularnewline
\hline 
\end{tabular}
\par\end{raggedright}
\begin{raggedright}
\textcolor{black}{}%
\begin{tabular}{@{\hspace*{1mm}}p{3cm}@{}|@{\hspace*{1mm}}p{2.8cm}|@{\hspace*{1mm}}p{9.3cm}}
\multicolumn{3}{>{\centering}p{15cm}}{\textbf{\textcolor{black}{B: Input parameters}}}\tabularnewline
\hline 
\textbf{\textcolor{black}{Symbol}} & \textbf{\textcolor{black}{Value}} & \textbf{Description}\tabularnewline
$A_{p}$ & $66\,\mathrm{[Hz]}$ & amplitude of Gaussian rate profile\tabularnewline
$\sigma_{p}$ & $10$ & width of Gaussian rate profile\tabularnewline
$c_{p}$ & $0.48\,\mathrm{[Hz]}$ & offset of Gaussian rate profile\tabularnewline
$s_{p}$ & $\left[25,75,\ldots,475\right]$ & set of possible values for the center of the Gaussian $\mu_{p}$\tabularnewline
$t_{\mu}$ & $100\,\mathrm{[ms]}$ & interval after which a new value for $\mu_{p}$ is drawn\tabularnewline
$N_{\mu}$ & $100$ & number of intervals $t_{\mu}$\tabularnewline
\hline 
\end{tabular}
\par\end{raggedright}
\caption{\textbf{Neuron and input parameters for simulation of network producing
bidirectional connections using the Clopath rule.} The values are
taken from \citep[Tab. 1 and appendix]{Clopath10_344}. The same values
are used for the performance measurements shown in Figures 9 and 10.\label{tab:neuron_params_cl_network-1}}
\end{table}

\newpage{}

\begin{table}[H]
\begin{raggedright}
\textcolor{black}{}%
\begin{tabular}{@{\hspace*{1mm}}p{2cm}@{}|@{\hspace*{1mm}}p{4.3cm}|@{\hspace*{1mm}}p{8.8cm}}
\hline 
\multicolumn{3}{c}{\textbf{A: Simulation parameters}}\tabularnewline
\hline 
\textbf{\textcolor{black}{Symbol}} & \textbf{\textcolor{black}{Value}} & \textbf{Description}\tabularnewline
\textcolor{black}{$T_{p}$} & $1000\,\mathrm{ms}$ & pattern duration\tabularnewline
$N_{\mathrm{rep}}$ & $100$ & number of pattern repetitions\tabularnewline
$N_{\mathrm{p}}$ & $200$ & number of input spike trains\tabularnewline
$f_{\mathrm{p}}$ & $10\,\mathrm{Hz}$ & input firing rate\tabularnewline
$w_{g_{E}}$ & $\left(18\,\sin\left(2\pi t\right)+4.8\right)\,\mathrm{nS}$ & weights to generate periodic excitatory conductance\tabularnewline
$w_{g_{I}}$ & $3\,\mathrm{nS}$ & weights to generate constant inhibitory conductance\tabularnewline
\hline 
\end{tabular}
\par\end{raggedright}
\begin{raggedright}
\textcolor{black}{}%
\begin{tabular}{@{\hspace*{1mm}}p{2cm}@{}|@{\hspace*{1mm}}p{4.3cm}|@{\hspace*{1mm}}p{8.8cm}}
\multicolumn{3}{c}{\textbf{B: Parameters of }\texttt{\textbf{pp\_cond\_exp\_mc\_urbanczik
(soma)}}}\tabularnewline
\hline 
\textbf{\textcolor{black}{Symbol}} & \textbf{\textcolor{black}{Value}} & \textbf{Description}\tabularnewline
$C_{m}$ & $300\,\mathrm{pF}$ & membrane capacitance\tabularnewline
$E_{L}$ & $-70\,\mathrm{mV}$ & leak reversal potential\tabularnewline
$g_{L}$ & $30.0\,\mathrm{nS}$ & leak conductance\tabularnewline
$E_{\mathrm{ex}}$ & $0.0\,\mathrm{mV}$ & exc. reversal potential\tabularnewline
$E_{\mathrm{in}}$ & $-75.0\,\mathrm{mV}$ & inh. reversal potential\tabularnewline
$\tau_{\mathrm{ex}}$ & $3.0\,\mathrm{ms}$ & rise time of the exc. synaptic alpha funct.\tabularnewline
$\tau_{\mathrm{in}}$ & $3.0\,\mathrm{ms}$ & rise time of the inh. synaptic alpha funct.\tabularnewline
$t_{\mathrm{ref}}$ & $3.0\,\mathrm{ms}$ & refractory time\tabularnewline
\hline 
\end{tabular}
\par\end{raggedright}
\begin{raggedright}
\textcolor{black}{}%
\begin{tabular}{@{\hspace*{1mm}}p{2cm}@{}|@{\hspace*{1mm}}p{4.3cm}|@{\hspace*{1mm}}p{8.8cm}}
\multicolumn{3}{c}{\textbf{C: Parameters of }\texttt{\textbf{pp\_cond\_exp\_mc\_urbanczik
(dendrite)}}}\tabularnewline
\hline 
\textbf{\textcolor{black}{Symbol}} & \textbf{\textcolor{black}{Value}} & \textbf{Description}\tabularnewline
$C_{m}$ & $300\,\mathrm{pF}$ & membrane capacitance\tabularnewline
$E_{L}$ & $-70\,\mathrm{mV}$ & leak reversal potential\tabularnewline
$g_{L}$ & $30.0\,\mathrm{nS}$ & leak conductance\tabularnewline
$\tau_{\mathrm{ex}}$ & $3.0\,\mathrm{ms}$ & rise time of the exc. synaptic alpha funct.\tabularnewline
$\tau_{\mathrm{in}}$ & $3.0\,\mathrm{ms}$ & rise time of the inh. synaptic alpha funct.\tabularnewline
$\phi\left(U\right)$ & $\frac{0.15\,\mathrm{kHz}}{1+\frac{1}{2}\exp\left(\frac{-55\,\mathrm{mV}-U}{3\,\mathrm{mV}}\right)}$ & rate function\tabularnewline
$g_{\mathrm{sp}}$ & $600.0\,\mathrm{nS}$ & coupling dendrite to soma\tabularnewline
\hline 
\end{tabular}
\par\end{raggedright}
\caption{\textbf{Parameters of the simulation of the learning experiment using
the Urbanczik-Senn rule.} The same values of the neuron parameters
are used for the performance measurements shown in Figures 11 and
12.\label{tab:params_us-1}}
\end{table}

\newpage{}

\textcolor{black}{}
\begin{table}[H]
\begin{raggedright}
\textcolor{black}{}%
\begin{tabular}{@{\hspace*{1mm}}p{3cm}@{}|@{\hspace*{1mm}}p{12.2cm}}
\hline 
\multicolumn{2}{>{\centering}m{15cm}}{\textbf{\textcolor{black}{A: Model summary}}}\tabularnewline
\hline 
\textbf{\textcolor{black}{Populations}} & \textcolor{black}{Three: excitatory, inhibitory, external input}\tabularnewline
\textbf{\textcolor{black}{Connectivity}} & \textcolor{black}{random with fixed indegree}\tabularnewline
\textbf{\textcolor{black}{Neuron model}} & adaptive exponential integrate-and-fire (aeif, Clopath)/

two-compartment Poisson (two-comp., Urbanczik-Senn)\tabularnewline
\textbf{\textcolor{black}{Plasticity}} & Clopath synapse/

Urbanczik-Senn synapse\tabularnewline
\textbf{\textcolor{black}{Input}} & \textcolor{black}{independent homogeneous Poisson spike trains}\tabularnewline
\textbf{\textcolor{black}{Measurements}} & \textcolor{black}{---}\tabularnewline
\hline 
\end{tabular}
\par\end{raggedright}
\begin{raggedright}
\textcolor{black}{}%
\begin{tabular}{@{\hspace*{1mm}}p{3cm}@{}|@{\hspace*{1mm}}p{2.8cm}@{}|@{\hspace*{1mm}}p{9.3cm}}
\multicolumn{3}{>{\centering}m{15cm}}{\textbf{\textcolor{black}{B: Populations}}}\tabularnewline
\hline 
\textbf{\textcolor{black}{Name}} & \textbf{\textcolor{black}{Elements}} & \textbf{\textcolor{black}{Population size}}\tabularnewline
\textcolor{black}{E} & aeif/two-comp. & $N_{E}=4N_{I}$\tabularnewline
I & aeif/two-comp. & $N_{I}$\tabularnewline
E\textsubscript{ext} & Poisson generator & $1$\tabularnewline
\hline 
\end{tabular}
\par\end{raggedright}
\begin{raggedright}
\textcolor{black}{}%
\begin{tabular}{@{\hspace*{1mm}}p{3cm}@{}|@{\hspace*{1mm}}p{2.8cm}@{}|@{\hspace*{1mm}}p{2.8cm}@{}|@{\hspace*{1mm}}p{3.7cm}@{}|@{\hspace*{1mm}}p{2.8cm}@{}}
\multicolumn{5}{>{\centering}p{15cm}}{\textbf{\textcolor{black}{C: Connectivity}}}\tabularnewline
\hline 
\textbf{Name} & \textbf{\textcolor{black}{Source}} & \textbf{Target} & \textbf{Pattern} & \textbf{Weight}\tabularnewline
Exc & E & E+I & fixed in-degree $C_{E}$ & $J_{\mathrm{ex}}$\tabularnewline
Inh & I & E+I & fixed in-degree $C_{I}$ & $J_{\mathrm{in}}$\tabularnewline
Ext & E\textsubscript{Ext} & E+I & all-to-all & $J$\tabularnewline
\hline 
\end{tabular}
\par\end{raggedright}
\begin{raggedright}
\textcolor{black}{}%
\begin{tabular}{@{\hspace*{1mm}}p{3cm}@{}|@{\hspace*{1mm}}p{12.2cm}}
\multicolumn{2}{>{\centering}m{15cm}}{\textbf{\textcolor{black}{D: Neurons}}}\tabularnewline
\hline 
\textbf{\textcolor{black}{Name}} & \texttt{aeif\_psc\_delta\_clopath}\tabularnewline
\textbf{\textcolor{black}{Type}} & adaptive exponential integrate-and-fire\tabularnewline
\textbf{Details} & see \citet{Clopath10_344}\tabularnewline
\textbf{Parameters} & see \prettyref{tab:neuron_params_cl_network-1}\tabularnewline
\textbf{\textcolor{black}{Name}} & \texttt{pp\_cond\_exp\_mc\_urbanczik}\tabularnewline
\textbf{\textcolor{black}{Type}} & two-compartment neuron with spike generation via inhomogeneous Poisson
process\tabularnewline
\textbf{Details} & see \citet{Urbanczik14}\tabularnewline
\textbf{Parameters} & see \prettyref{tab:params_us-1}\tabularnewline
\hline 
\multicolumn{2}{c}{\textbf{E: Synapses}}\tabularnewline
\hline 
\textbf{Name} & \textbf{Model}\tabularnewline
Exc & \texttt{clopath/urbanczik\_synapse}\tabularnewline
Inh & \texttt{clopath/urbanczik\_synapse}\tabularnewline
Ext & \texttt{static\_synapse}\tabularnewline
\hline 
\end{tabular}
\par\end{raggedright}
\begin{raggedright}
\textcolor{black}{}%
\begin{tabular}{@{\hspace*{1mm}}p{3cm}@{}|@{\hspace*{1mm}}p{12.2cm}}
\multicolumn{2}{>{\centering}m{15cm}}{\textbf{\textcolor{black}{F: Input}}}\tabularnewline
\hline 
\textbf{\textcolor{black}{Type}} & \textbf{\textcolor{black}{Description}}\tabularnewline
\textcolor{black}{Poisson generator} & \textcolor{black}{homogeneous Poisson spike trains, independent for
each neuron, rate $f_{\mathrm{ext}}=\nuext C_{E}$}\tabularnewline
\hline 
\end{tabular}
\par\end{raggedright}
\raggedright{}\textcolor{black}{\caption{\textbf{Model description of Brunel network after }\citet{Nordlie-2009_e1000456}.
This network is used to produce the performance measurement shown
in Figures 9, 10, 11, and 12. The values of the parameters are shown
in \prettyref{tab:Parameters_brunel}. \label{tab:Network-and-simulation}}
}
\end{table}

\newpage{}

\textcolor{black}{}
\begin{table}[H]
\begin{raggedright}
\textcolor{black}{}%
\begin{tabular}{@{\hspace*{1mm}}p{3cm}@{}|@{\hspace*{1mm}}p{2.8cm}@{}|@{\hspace*{1mm}}p{9.4cm}}
\hline 
\multicolumn{3}{>{\centering}m{15cm}}{\textbf{\textcolor{black}{A: Global simulation parameters}}}\tabularnewline
\hline 
\textbf{\textcolor{black}{Symbol}} & \textbf{Value} & \textbf{Description}\tabularnewline
$T$ & $2\cdot10^{3}\,\mathrm{ms}$ & biological time\tabularnewline
$h$ & $0.1\,\mathrm{ms}$ & resolution\tabularnewline
\hline 
\end{tabular}
\par\end{raggedright}
\begin{raggedright}
\textcolor{black}{}%
\begin{tabular}{@{\hspace*{1mm}}p{3cm}@{}|@{\hspace*{1mm}}p{2.8cm}@{}|@{\hspace*{1mm}}p{9.4cm}}
\multicolumn{3}{>{\centering}m{15cm}}{\textbf{\textcolor{black}{B: Network sizes}}}\tabularnewline
\hline 
\textbf{\textcolor{black}{Symbol}} & \textbf{\textcolor{black}{Value}} & \textbf{\textcolor{black}{Description}}\tabularnewline
\textcolor{black}{$N=N_{E}+N_{I}$} & $1.92\cdot10^{6}$ & number of neurons in Clopath simulation with small indegree $K=100$\tabularnewline
$N$ & $1.54\cdot10^{5}$ & number of neurons in Clopath simulation with large indegree $\mbox{\ensuremath{K=5000}}$\tabularnewline
$N$ & $3.84\cdot10^{5}$ & number of neurons in Urbanczik simulation with small indegree $K=100$\tabularnewline
$N$ & $3.84\cdot10^{4}$ & number of neurons in Urbanczik simulation with large indegree $K=5000$\tabularnewline
\hline 
\end{tabular}
\par\end{raggedright}
\begin{raggedright}
\textcolor{black}{}%
\begin{tabular}{@{\hspace*{1mm}}p{3cm}@{}|@{\hspace*{1mm}}p{2.8cm}@{}|@{\hspace*{1mm}}p{9.4cm}}
\multicolumn{3}{>{\centering}m{15cm}}{\textbf{\textcolor{black}{C: Connectivity}}}\tabularnewline
\hline 
\textbf{Symbol} & \textbf{\textcolor{black}{Value}} & \textbf{Description}\tabularnewline
$g$ & $7.0$ & ratio inh./exc. weight\tabularnewline
$J$ & $0.1$ & postsynaptic amplitude

The unit depends on the neuron model. In case of the aeif model and
the Clopath rule it is $[\mathrm{mV}]$ and for the Urbanczik-Senn
rule it is $\mathrm{[pA]}$\tabularnewline
$J_{\mathrm{ex}}$ & $J$ & amplitude of exc. postsyn. potential\tabularnewline
$J_{\mathrm{in}}$ & $-gJ_{\mathrm{ex}}$ & amplitude of inh. postsyn. potential\tabularnewline
$K=C_{E}+C_{I}$ & $100$ or $5000$ & total number of excitatory synapses per neuron\tabularnewline
$C_{E}$ & $0.8K$ & number of excitatory synapses per neuron\tabularnewline
$C_{I}$ & $0.2K$ & number of inhibitory synapses per neuron\tabularnewline
$\eta$ & $0.0$ & learning rate\tabularnewline
\hline 
\end{tabular}
\par\end{raggedright}
\begin{raggedright}
\textcolor{black}{}%
\begin{tabular}{@{\hspace*{1mm}}p{3cm}@{}|@{\hspace*{1mm}}p{2.8cm}@{}|@{\hspace*{1mm}}p{9.4cm}}
\multicolumn{3}{>{\centering}m{15cm}}{\textbf{\textcolor{black}{D: External input}}}\tabularnewline
\hline 
\textbf{\textcolor{black}{Symbol}} & \textbf{Value} & \textbf{Description}\tabularnewline
\textbf{\textcolor{black}{$\nuext$}} & $6.75\cdot10^{-3}\,\mathrm{Hz}$ & factor in rate of external Poisson input $f_{\mathrm{ext}}=\nuext C_{E}$\tabularnewline
\hline 
\end{tabular}
\par\end{raggedright}
\raggedright{}\textcolor{black}{\caption{\textbf{Parameters of the Brunel network}.\label{tab:Parameters_brunel}}
}
\end{table}

\end{document}